\documentclass[aps,pra,reprint,amsmath,amssymb,superscriptaddress,floatfix]{revtex4-1}

\usepackage{natbib} 
\usepackage[colorlinks,
            linkcolor=blue,
            anchorcolor=blue,
            citecolor=blue,
urlcolor=blue
            ]{hyperref}
\usepackage{graphicx}
\usepackage[FIGTOPCAP,raggedright,nooneline]{subfigure}
\usepackage{braket}
\usepackage{mathtools}
\usepackage{float}

\begin{document}


\title{Adaptive tomography of qubits: Purity versus statistical fluctuations}


\author{Aonan Zhang}
\email[]{aonan.zhang@smail.nju.edu.cn}
\affiliation{National Laboratory of Solid State Microstructures and College of Engineering and Applied Sciences, Nanjing University, Nanjing 210093, China}
\affiliation{Collaborative Innovation Center of Advanced Microstructures, Nanjing University, Nanjing 210093, China}
\author{Yujie Zhang}
\affiliation{Kuang Yaming Honors School, Nanjing University, Nanjing 210093, China}
\author{Feixiang Xu}
\affiliation{National Laboratory of Solid State Microstructures and College of Engineering and Applied Sciences, Nanjing University, Nanjing 210093, China}
\affiliation{Collaborative Innovation Center of Advanced Microstructures, Nanjing University, Nanjing 210093, China}
\author{Long Li}
\affiliation{Collaborative Innovation Center of Advanced Microstructures, Nanjing University, Nanjing 210093, China}
\affiliation{School of Physics, Nanjing University, Nanjing 210093, China}
\author{Lijian Zhang}
\email[]{lijian.zhang@nju.edu.cn}
\affiliation{National Laboratory of Solid State Microstructures and College of Engineering and Applied Sciences, Nanjing University, Nanjing 210093, China}
\affiliation{Collaborative Innovation Center of Advanced Microstructures, Nanjing University, Nanjing 210093, China}


\date{\today}

\begin{abstract}
The success of quantum information processing applications relies on accurate and efficient characterization of quantum states, especially nearly-pure states. In this work, we investigate a procedure for adaptive qubit state tomography which achieves $O(1/N)$ scaling in accuracy for large $N$. We analyze the performance of the adaptive protocol on the characterization of pure, mixed and nearly-pure states, and clarify the interplay between the purity of the state and the statistical fluctuations of the measurement results. Our results highlight the subtle difference between the characterization of nearly-pure states and those of pure or highly mixed states.
\end{abstract}

\keywords{Quantum tomography; quantum information processing; quantum measurement}

\maketitle

\section{\label{sec:1}Introduction}
Implementation of quantum information processing (QIP) relies on accurate characterization and manipulation of quantum states. Therefore the estimation of an unknown quantum state is of fundamental importance in quantum information theory  \cite{RN1,RN2}. Quantum state tomography (QST) is a process of identifying the density matrix $\rho$ of an unknown quantum state via measurements on $N$ identical copies of the state. Due to the statistical fluctuations of the measurement results, there would be a statistical distance between the estimate $\hat{\rho}$ and the true state $\rho$. Tomographic protocols aim at minimizing the statistical distance with limited resources \cite{RN3,RN4,RN5}, while a variety of metrics have been used to quantify the statistical distance \cite{RN6}. One of the well-motivated metrics is the \textit{infidelity}, which is defined as

\begin{equation}
1-F(\rho,\hat{\rho})=1-\mathrm{Tr}\left(\sqrt{\sqrt{\rho}\hat{\rho}\sqrt{\rho}}\right)^2.
\label{eq:1}
\end{equation}
\par
Strictly speaking, estimating the ``\textit{unknown state}'' means that there is no priori information, i.e., the probability distribution for all possible states within the Hilbert space is homogeneous before the estimation procedure. In practice, priori knowledge about the quantum state can be given in various ways, such as that the state is pure, or that the priori probability distribution of the state is not uniform \cite{RN7,RN8}. Such priori information may help to design accurate and efficient QST protocols.
\par
How to gain priori knowledge about an unknown state? One simple idea is to perform pre-measurements of the state and acquire a partial characterization. Such procedure is known as the adaptive quantum state tomography, which has been investigated in a host of theoretical and experimental works \cite{RN11,RN12,RN13,RN14,RN15,RN16,RN17,arxiv1604,PhysRevA.85.052107}. Adaptive tomography has the advantages of improving the characterization accuracy and being less sensitive to systematic errors than standard tomography. Recently, a simple protocol for adaptive tomography based on mutually unbiased bases (MUB) has been proposed \cite{RN12}, and improves the estimation accuracy, which is quantified by the infidelity, from $O(1/\sqrt{N})$ to $O(1/N)$ for pure states. Furthermore, adaptive tomography using Bayesian  estimation has been demonstrated and achieves $O(1/N)$ scaling in the estimation accuracy for pure states \cite{RN13}. 
\par
However, in practical setups the noise is unavoidable in the preparation of a quantum state and measurement apparatus, resulting in nearly-pure states rather than pure states. To date, the performance of adaptive tomography on the indispensable set of quantum states for QIP, the nearly-pure states, remains to be elucidated even for the single-qubit scenario. In particular, how the purity of the state affects the performance has not been clarified entirely. In this work, we study the adaptive quantum state tomography using two quintessential measurements: the measurement with mutually unbiased bases (MUB) and the symmetric informationally complete positive operator-valued measure (SIC-POVM), and analyze their performance on pure, highly mixed and nearly-pure states. It is shown that the interplay between the purity of the state and the statistical fluctuations affects the scaling of the estimation accuracy, and defines the transition region from pure to mixed states, i.e., the nearly-pure states.
\par
The rest of the paper is organized as follows. In Sec. \ref{sec:2} adaptive quantum state tomography using MUB and SIC-POVM are introduced, including the optimal configuration, the maximum likelihood estimation (MLE), and the adaptive protocols. Then we analyze the performance of QST for pure and highly mixed states.  Special attention is paid to nearly-pure states in Sec. \ref{sec:4}, in which the purity of the state and the statistical fluctuations in the measurement results are comparable and affect the performance of tomography. Finally, we discuss the criterion of nearly-pure states in QST.

\section{\label{sec:2}Adaptive quantum state tomography}
\subsection{Tomography using MUB and SIC-POVM}
\par
A quantum state reveals its information through measurements \cite{RN5}. Optimal measurement sets for QST based on various figure of merits have been investigated \cite{RN18,RN19,RN20,RN21,RN32,RN40}. Mutually unbiased bases (MUB) are optimal fixed measurements for von Neumann measurements \cite{RN22} and widely used in standard state tomography. Two orthogonal bases $\{|u_1\rangle,...,|u_d\rangle\}$ and $\{|v_1\rangle,...,|v_d\rangle\}$ are mutually unbiased if
\begin{equation}
|\langle u_i|v_j\rangle|^2=1/d \quad \forall i,j.
\end{equation}
We should indicate that the MUB measurements are projective measurements. For a $d$-dimensional system, a quantum state is specified by $d^2-1$ real parameters, therefore MUB measurements need at least $(d^2-1)/(d-1)=d+1$ orthogonal measurements. If $d=2$, MUB measurements require 3 orthogonal measurements to identify the density matrix:
\begin{equation}
\rho=\frac{1}{2}\left(\openone+\vec{s}\cdot\vec{\sigma}\right),
\label{eq:2}
\end{equation}
where $\vec{s}=(s_x,s_y,s_z)$ is the Pauli vector and $\vec{\sigma}=(\sigma_x,\sigma_y,\sigma_z)$ is the tensor of Pauli operators. Complete sets of MUB have been constructed in Hilbert spaces whose dimensions are any power of prime, but not scalable to arbitrary high dimensions \cite{RN23,RN24}, even as low as $d=6$ \cite{MCNULTY_2012}.
\begin{figure}
\centering
\addtocounter{figure}{1}
\subfigure[]{\includegraphics[width=0.45\linewidth]{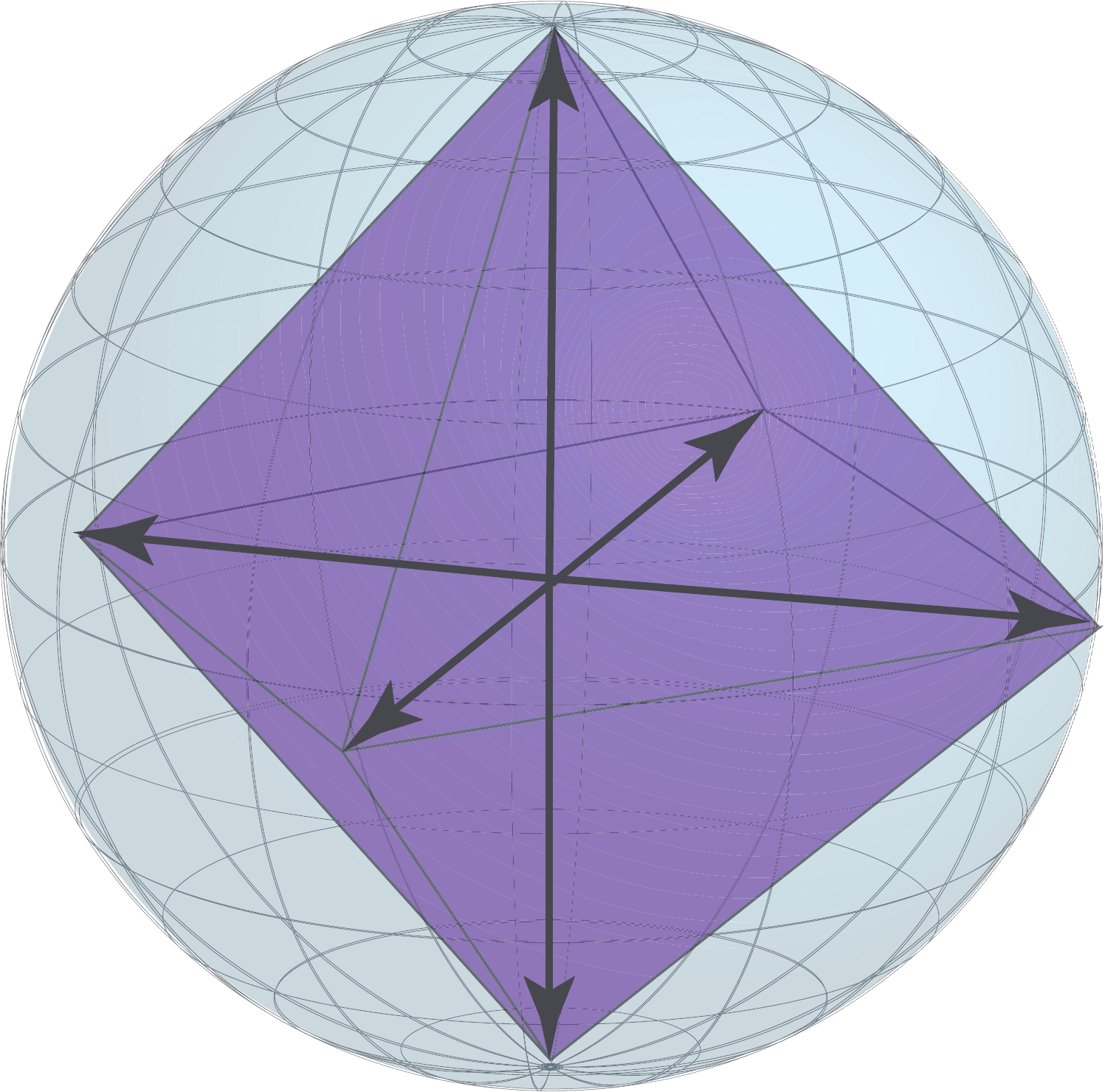}\label{fig:1:a}}\quad
\subfigure[]{\includegraphics[width=0.45\linewidth]{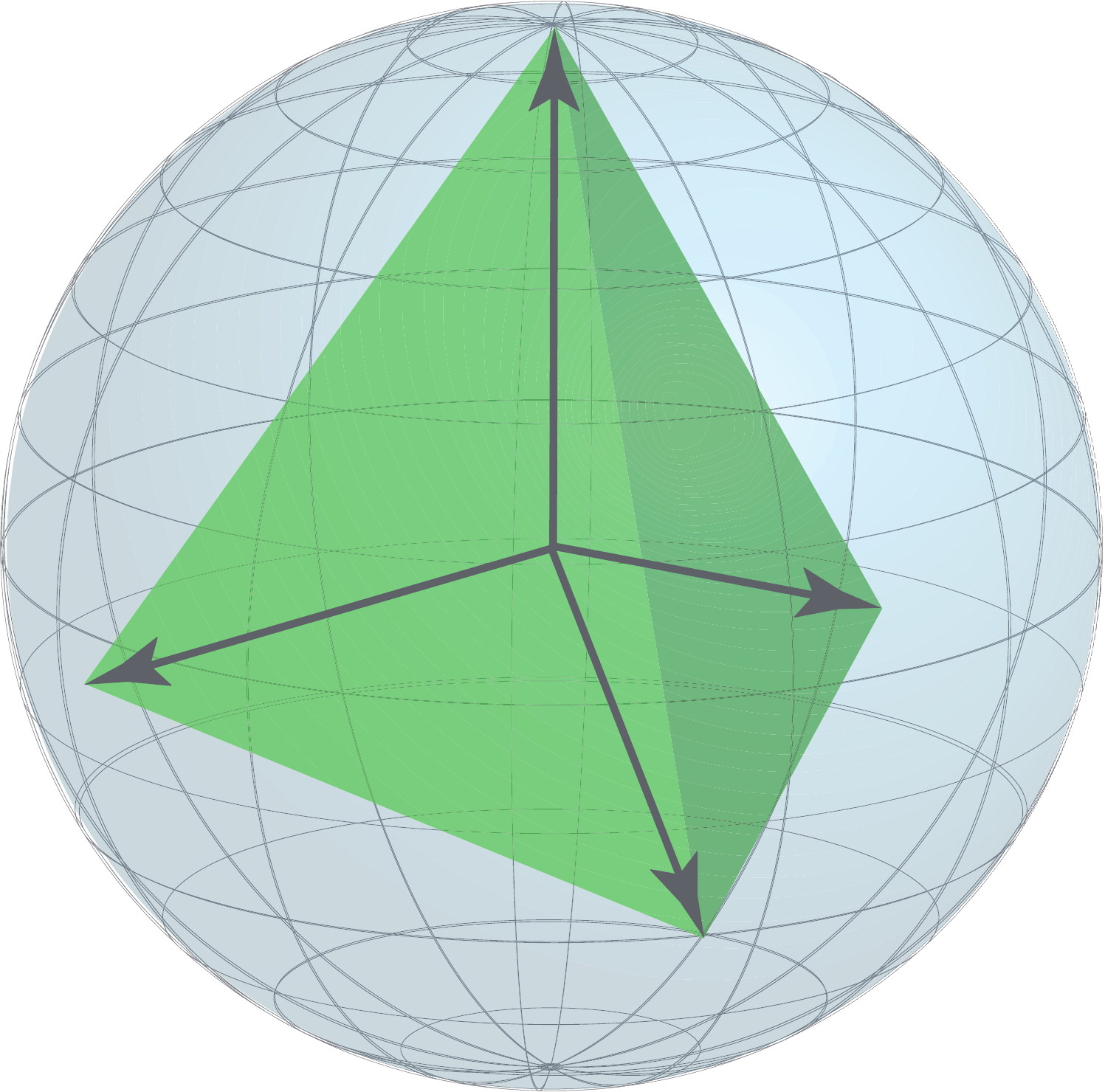}\label{fig:1:b}}
\addtocounter{figure}{-1}
 \caption{\label{fig:1}Projections of MUB and SIC-POVM in Bloch sphere($d=2$). (a) The six measurement outcomes of MUB constitute the vertices of an octahedron at Bloch sphere. (b) The four measurement outcomes of SIC-POVM constitute the vertices of a tetrahedron.}
\end{figure}
\par
On the other side, positive-operator valued measures (POVM) allow one to gain more information from a single measurement than the standard projective measurements. In quantum measurement theory, a positive-operator-valued measure (POVM) can be defined as a set of non-negative operators $\{E_i\}$ which fulfills the completeness condition $\sum_{i}E_i=\openone$. The POVM is informationally complete when every state is determined completely by the measurement statistics, and more useful when the POVM is symmetric when all pairwise inner products between the POVM elements are equal \cite{RN23,RN25}. In mathematical viewpoint, MUB and SIC-POVM for a qubit are both measurement sets as platonic solids in Bloch sphere \cite{RN32,RN33}, as shown in Fig. \ref{fig:1}.

\par
SIC measurements are of vital importance in quantum information theory, especially in quantum tomography \cite{RN26,RN34,RN35,RN28} and quantum communication \cite{PhysRevA.83.062304,1751-8121-47-44-445301,1751-8121-47-23-235302,PhysRevA.94.022335}. State tomography with SIC-POVM has been experimentally demonstrated for qubits \cite{RN34}, qutrits \cite{RN35} and qudits \cite{RN28}. In contrast to the MUB, SIC-POVM minimizes the amount of measurements, and has been conjectured to exist in arbitrary dimensions \cite{RN25,RN27,RN28,RN29}. SIC-POVM can be implemented with a single measurement setting \cite{RN9,RN10,RN30,RN31}, while MUB usually changes the configuration to realize multiple measurement sets and divides the ensemble of states into subgroups. In the following, we will focus on the SIC-POVM because previous works have not given a detailed analysis of adaptive tomography using SIC-POVM, while we also give the results for MUB. For a $d$-dimensional quantum system, SIC-POVM is described by a set of projectors $P_i=\ket{\psi_i}\bra{\psi_i} \quad (i=0,1,2...d^2-1)$, which satisfy
\begin{equation}
\mathrm{Tr}\left(P_i P_j\right)=\frac{1}{d+1} \ (\text{for}\ i\neq j), \quad E_i=\frac{1}{d}P_i.
\label{eq:3}
\end{equation}

\par
The projectors $P_i$  at Bloch sphere are sketched in Fig. \ref{fig:1:b}. Attributed to the constraint of completeness, all possible measurement results are limited to the interior of the tangential tetrahedron of the Bloch sphere. Geometrically, the four tangent points are opposite to the four projectors. In accordance with Born's rule, the probability of observing outcome $i$ is $p_i=\mathrm{Tr}(E_i \rho)$, which can be represented by the Pauli parameters
\begin{eqnarray}
p_0=&&\frac{1}{4}\left(1+s_z\right), \nonumber \\
p_1=&&\frac{1}{4}\left(1+\frac{2\sqrt{2}}{3}s_x-\frac{1}{3}s_z\right), \nonumber \\
p_2=&&\frac{1}{4}\left(1-\frac{\sqrt{2}}{3}s_x+\frac{\sqrt{6}}{3}s_y-\frac{1}{3}s_z\right), \nonumber \\
p_3=&&\frac{1}{4}\left(1-\frac{\sqrt{2}}{3}s_x-\frac{\sqrt{6}}{3}s_y-\frac{1}{3}s_z\right).
\label{eq:5}
\end{eqnarray}
The density matrix $\rho$ can be parameterized through probabilities $\vec{p}=(p_0,p_1,p_2,p_3)$ as $\rho=6\vec{p}\cdot\vec{E}-1$, which yields
\begin{widetext}
\begin{equation}
\rho=
\begin{pmatrix}
2p_0 & \sqrt{2}\left(p_1+p_2\mathrm{e}^{-\mathrm{i}2\pi/3}+p_3\mathrm{e}^{\mathrm{i}2\pi/3}\right) \\
\sqrt{2}\left(p_1+p_2\mathrm{e}^{\mathrm{i}2\pi/3}+p_3\mathrm{e}^{-\mathrm{i}2\pi/3}\right) & 1-2p_0
\end{pmatrix}
.
\label{eq:6}
\end{equation}
\end{widetext}
\subsection{\label{sec:accuracy}Tomography accuracy and optimal strategy}
\par
In the single-qubit tomography scenario, the outcomes of SIC-POVM subject to a multinomial distribution $\mathcal{M}(N,(p_0,p_1,p_2,p_3))$, of which the probability mass function is
\begin{equation}
f(n_i;N;p_i)=\frac{N}{\prod_i n_i !}\prod_i p^{n_i}_i,
\label{eq:multi}
\end{equation}
If we focus on the estimation of a specific parameter $p_i$, $n_i$ becomes a Bernoulli trial $\sim B(N,p_i)$ with a binomial distribution. The expectation and standard deviation of estimator $\hat{p}_i$ are
\begin{equation}
\langle\hat{p}_i\rangle=p_i,\quad \Delta\hat{p}_i=\sqrt{\frac{1}{N}}\sqrt{p_{i}(1-p_{i})}.
\label{eq:13}
\end{equation}
The estimated state is subject to such statistical fluctuations and the physical constraints. As an example, for the parameter $p_0$ in Eq. (\ref{eq:5}), we can substitute $s_z$ [Eq. (\ref{eq:5})] into Eq. (\ref{eq:13}), and yield
\begin{equation}
\langle\hat{s}_z\rangle=s_z,\quad \Delta\hat{s}_z=\sqrt{\frac{4}{N}}\sqrt{1-{\left(\frac{1-s_z}{2}\right)}^2}.
\label{eq:14}
\end{equation}
In general cases, $\Delta\hat{s}_x,\Delta\hat{s}_y$ are $O(1/\sqrt{N})$ as well. Meanwhile, the infidelity defined in Eq. (\ref{eq:1}) can also be given in terms of Pauli vector representation \cite{RN39}
\begin{equation}
1-F(\vec{s},\vec{s}_{\mathrm{est}})=\frac{1}{2}\left(1-\vec{s}\cdot\vec{s}_{\mathrm{est}}-\sqrt{1-s^2}\sqrt{1-{s_{\mathrm{est}}}^2}\right),
\label{eq:10}
\end{equation}
where $\vec{s}_\mathrm{est}$ is the Pauli vector of the estimate state and $s=| \vec{s}|$ ($s_{\mathrm{est}}=|\vec{s}_{\mathrm{est}}|$) denotes the length of $\vec{s}$ ($\vec{s}_{\mathrm{est}}$). The discrepancy between the true state $\vec{s}$ and the estimate state $\vec{s}_{\mathrm{est}}$ is denoted as $\delta \vec{s}=\vec{s}-\vec{s}_{\mathrm{est}}$, and $|\delta\vec{s}|\propto O(1/\sqrt{N})$ since $\Delta\hat{s}_x,\Delta\hat{s}_y,\Delta\hat{s}_z\propto O(1/\sqrt{N})$. For a typical mixed state, the average infidelity can be estimated under the approximation $|\delta\vec{s}|\ll 1-s$  \cite{RN41}
\begin{equation}
1-F\approx\frac{1}{4}{|\delta\vec{s}|}^2\propto O(1/N).
\label{eq:11}
\end{equation}
However, if the true state is (or approaches) a general pure state, $| \vec{s}|$ equals (or approaches) 1. The third term on the right side of Eq. (\ref{eq:10}) vanishes, and the infidelity can be inferred by
\begin{eqnarray}
1-F=&&\frac{1}{2}\left(1-\vec{s}\cdot\vec{s}_{\mathrm{est}}\right) \nonumber \\
=&&\frac{1}{2}\left(1-\vec{s}\cdot\vec{s}\right)+\frac{1}{2}\vec{s}\cdot\delta\vec{s}\propto O(1/\sqrt{N}).
\label{eq:12}
\end{eqnarray}
The $O(1/\sqrt{N})$ term dominates in the average of Eq. (\ref{eq:12}) over the distribution of the measurement results, therefore the average infidelity scales as $O(1/\sqrt{N})$, as shown in Fig. \ref{fig:estimation:a}. However, not all of the pure states behave as $O(1/\sqrt{N})$ scaling. When the measurement is aligned in several specific configurations, the tomography accuracy would be significantly improved. Such configurations are called the optimal strategy of QST. The optimal strategy for MUB has been investigated in previous works \cite{RN12}. If one basis diagonalizes the true state $\rho$, the infidelity would be minimized. For SIC-POVM, the optimal strategy is the antiparallel strategy from both the tomographic purpose~\cite{RN37} and the viewpoint of informational power~\cite{PhysRevA.83.062304,QIP15.2016}. In Eq. (\ref{eq:14}), when $s_z\approx -1$ (i.e., $p_0\approx 0$), the uncertainty in the estimation $\hat{s}_z$ (i.e., $\hat{p_0}$) is reduced to 0. The true state is aligned antiparallel to $\ket{\psi_0}$ in Bloch sphere. Similarly, if the state is antiparallel to any $\ket{\psi_i}$, we will have similar results.
\begin{figure}
\centering
\addtocounter{figure}{1}
\subfigure[]{\includegraphics[width=0.45\linewidth]{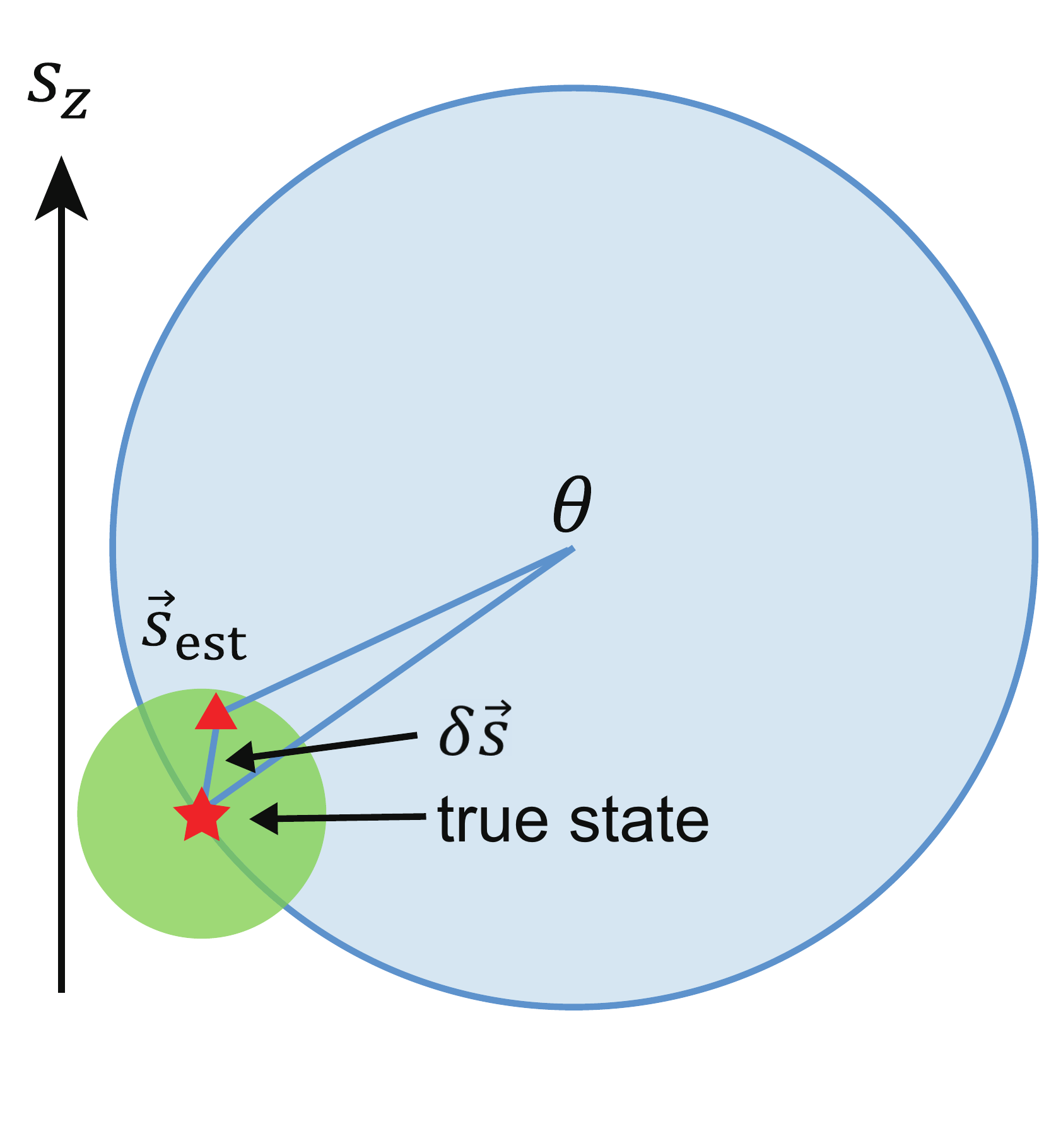}\label{fig:estimation:a}}\quad
\subfigure[]{\includegraphics[width=0.45\linewidth]{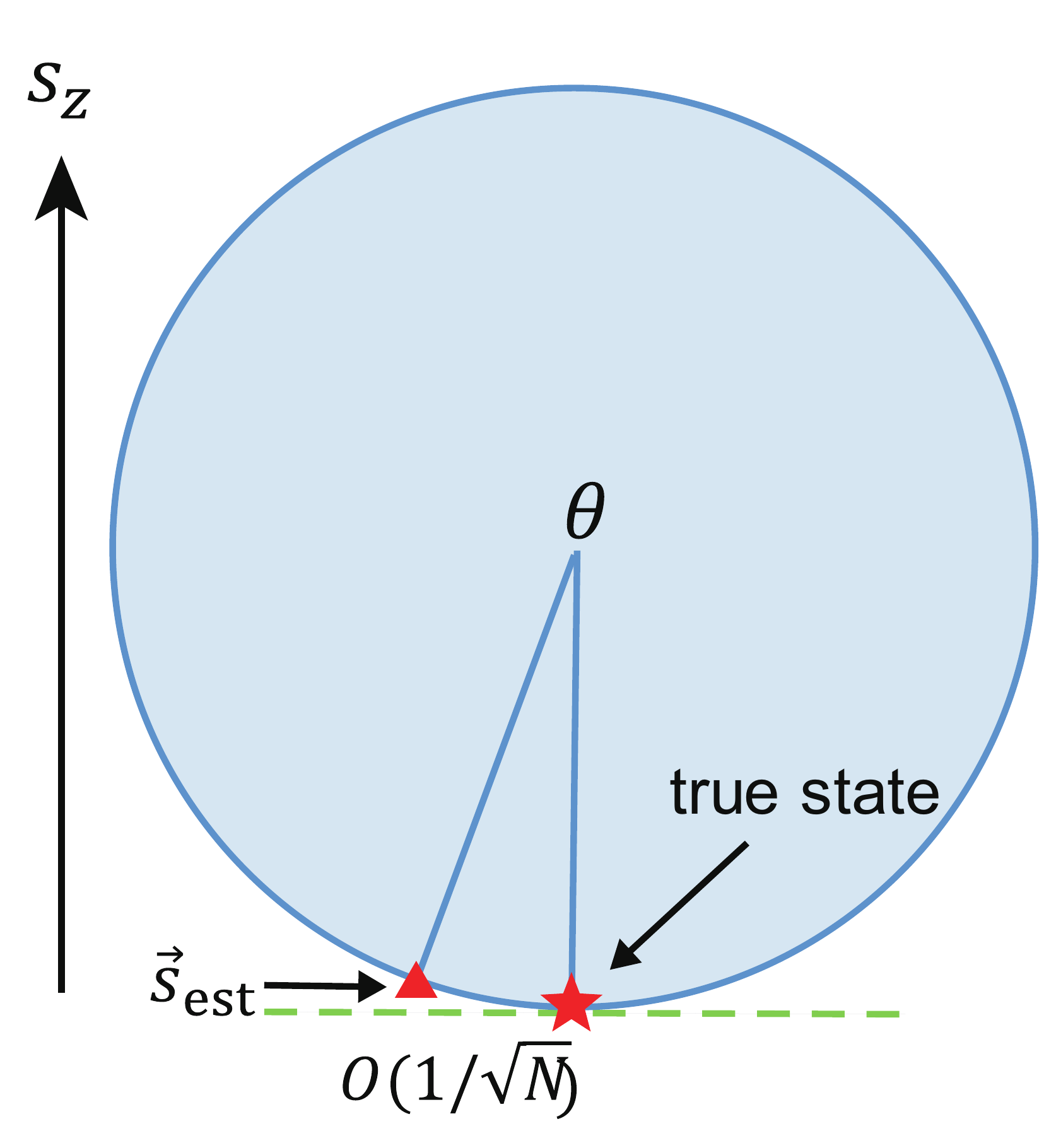}\label{fig:estimation:b}}
\addtocounter{figure}{-1}
 \caption{\label{fig:estimation}The scenario of estimating pure states. Each figure represents the cross-section of Bloch sphere. The red star denotes the true state $\vec{s}$, and the triangle denotes the estimation $\vec{s}_{\mathrm{est}}$. (a) Estimation of a general pure state. The average infidelity is obtained by averaging the infidelity over the distribution of the measurement results (shaded area, green). (b) Estimation of a pure state in the optimal configuration. The probability of observing one of the outcomes becomes 0, therefore the distribution of the measurement results is a planar region (dashed line, green).}
\end{figure}
\par
Figure \ref{fig:estimation:b} illustrates the scenario of estimating a pure state in the optimal configuration. In this case, one of the outcomes is not registered at all; the distribution of the measurement results is compressed into a plane without statistical error in $s_z$ direction. The statistical fluctuations on $s_x$ and $s_y$ are $O(1/\sqrt{N})$ . Meanwhile, because most of trials violate the physical constraint, the estimation $\vec{s}_\mathrm{est}$, as well as the true state $\vec{s}$, is a pure state. In consequence, we can estimate the infidelity
\begin{eqnarray}
1-F=&&\frac{1}{2}\left(1-\vec{s}\cdot\vec{s}_{\mathrm{est}}\right) \nonumber \\
\approx&&\frac{1}{2}\left(1-\cos{\theta}\right)\propto O(1/N).
\end{eqnarray}
The angle $\theta\approx |\delta\vec{s}|/s\propto O(1/\sqrt{N})$ indicates the discrepancy between $\vec{s}_\mathrm{est}$ and $\vec{s}$. Back to Eq. (\ref{eq:12}), the projection between $\vec{s}$ and the direction vector $\delta\vec{s}/|\delta\vec{s}|$, as well as $|\delta\vec{s}|$ itself, is $O(1/\sqrt{N})$, therefore the inner product $\vec{s}\cdot\delta\vec{s}$ in Eq. (\ref{eq:12}) leads to an infidelity scaling as $O(1/N)$. 

\subsection{\label{sec:mle}Maximum likelihood estimation}
\par
In actual experiments, we collect the count $n_i$ of the measurement outcome $i$ rather than the well-defined probability $p_i$. Accordingly we substitute empirical frequencies $\hat{p}_i$ that is statistically determined by $p_i$ and reconstruct the density matrix. Due to the statistical fluctuations of the measurement values, it is possible to violate the positivity constraint if we directly apply a linear inversion with the raw data [Eq. (\ref{eq:6})], especially for pure states. In practical tomography, maximum likelihood estimation (MLE) is used extensively to reconstruct a physical state \cite{RN4,RN36}. In compliance with the Bayesian principle of statistical inference, MLE maximizes the likelihood function \cite{RN37,RN38}
\begin{equation}
\mathcal{L}\left(n_i|\rho\right)=\prod_{i}\left[\mathrm{Tr}\left(E_i\rho\right)\right]^{n_i}
\label{eq:7}
\end{equation}
subjects to the conditions $\rho\geq0$ and $\mathrm{Tr}(\rho)=1$. Generally, this problem can be conducted using standard tools for convex optimization. Nevertheless, the extremum algorithm proposed in \cite{RN37} is instructive to give an analytical derivation. Here we take the scenario of SIC-POVM as an example to explain the mechanism of MLE. By incorporating the likelihood function with Lagrange multipliers, the extremum equations can be written of the form
\begin{gather}
\mu+2-\frac{1}{2}\sum_i\sqrt{{1-\mu}^2+12\mu\hat{p}_i}=0,\label{eq:8} \\
\frac{\hat{p}_i}{\tilde{p}_i}=1-\mu+3\mu\tilde{p}_i,\label{eq:9}
\end{gather}
where $\mu$ is a Lagrange multiplier, $\tilde{p}_i$ denotes the probabilities after MLE applied. For a pure state that is antiparallel to one of the projections in SIC-POVM, for example $|\psi_0\rangle$, Eqns. (\ref{eq:8}) and (\ref{eq:9}) are solved near $\vec{p}=(0,1/3,1/3,1/3)$ and give the solution
\begin{equation}
\tilde{p}_i=\sqrt{\frac{1}{3}\hat{p}_i}.
\label{eq:19}
\end{equation}
According to the multinomial distribution, the measurement results of the outcomes are
\begin{gather}
\hat{p}_0= 0;\nonumber \\
\langle\hat{p}_i\rangle=\frac{1}{3},\quad\Delta\hat{p}_i=\sqrt{\frac{2}{9N}}\quad (i=1,2,3).
\label{eq:20}
\end{gather}
By means of Eq. (\ref{eq:19}), we get the probabilities $\tilde{p}_i$ after MLE applied under large $N$ approximation
\begin{gather}
\tilde{p}_0=1-\tilde{p}_1-\tilde{p}_2-\tilde{p}_3;\nonumber \\
\langle\tilde{p}_i\rangle=\frac{1}{3},\quad\Delta\tilde{p}_i\approx\frac{1}{2}\sqrt{\frac{2}{9N}}\quad (i=1,2,3).
\label{eq:21}
\end{gather}
Consequently, MLE not only compresses the estimation $\vec{s}_{\mathrm{est}}$ to a physical state, but also reduces the statistical error of $\hat{p}_i$ by a factor of 2. Then we define the effective length of the state
\begin{equation}
s_\mathrm{eff}=\sqrt{{\langle{\hat{s}}\rangle}^2+\langle{|\delta\vec{s}|}^2\rangle}=\sqrt{12\sum_i\langle\hat{p}_i^2\rangle-3}
\label{eq:eff}
\end{equation}
to represent the typical length of the reconstructed Pauli vector. Following the multinomial distribution Eq. (\ref{eq:multi}), one can derive the sum of $\langle \hat{p}_i^2\rangle$:
\begin{eqnarray}
\sum_i \langle \hat{p}_i^2\rangle=&&\frac{1}{N}+\frac{N-1}{N}\sum_i p_i^2 \nonumber \\
=&&\frac{3+s^2}{12}+\frac{9-s^2}{12N},
\end{eqnarray}
where $\sum_i p_i^2 =(3+s^2)/12$. Therefore, Eq. (\ref{eq:eff}) can be represented as $s_\mathrm{eff}=\sqrt{s^2+(9-s^2)/N}$ in general cases. In the estimation of a pure state in the optimal configuration, $\hat{s}$ is always greater or equal to 1, and MLE is scarcely avoidable, as shown in Fig. \ref{fig:estimation:b}. In virtue of the compression effect of MLE, the discrepancy angle $\theta\approx|\delta\vec{s}|=\sqrt{(9-s^2)/4N}$. To see the MLE of the measurement result is denoted by $\vec{s}_\mathrm{mle}$, it suffices to note that
\begin{equation}
1-F=\frac{1}{2}\left(1-\vec{s}\cdot\vec{s}_\mathrm{mle}\right)\approx\frac{1}{4}\theta^2\approx\frac{1}{2N}.
\label{eq:22}
\end{equation}
\par
For MUB, we can utilize similar analysis and obtain $\theta\approx|\delta\vec{s}|=\sqrt{(9-3s^2)/N}$ in general cases. The infidelity for a pure state in the optimal configuration can be estimated as
\begin{equation}
1-F=\frac{1}{2}\left(1-\vec{s}\cdot\vec{s}_\mathrm{mle}\right)\approx\frac{2}{3N}.
\label{eq:MUBinf}
\end{equation}

\subsection{\label{sec:adaptive}Adaptive protocol and simulation results}

\par
From the above analysis, it is shown that the behavior of QST is basis dependent. However, to approach the optimal configuration one requires some priori knowledge about the state. To achieve this goal, a two-step adaptive protocol can be employed: firstly, perform pre-estimation on half of copies $N_1=N/2$ of the state through static tomography and acquire an estimation $\hat{\rho}_0$; secondly, transform the measurement to the optimal configuration with respect to $\hat{\rho}_0$, and obtain a more accurate estimation of the state through the remaining $N/2$ copies. The two-step adaptive protocol allows us to approach the optimal configuration in the second step with the information acquired on the state within the first step. This procedure is expected to increase the accuracy prominently, resulting in $O(1/N)$ scaling for pure states. The adaptive protocol using MUB has been demonstrated for single qubits \cite{RN12}. For single qubit, arbitrary set of rank 1 and rank 2 POVM elements including SIC-POVM can be implemented by a one-dimensional quantum walk \cite{RN30}, which suggests that the adaptive protocol using SIC-POVM is also feasible in experimental implementation.

\begin{figure*}
\centering
\addtocounter{figure}{1}
\subfigure[]{\includegraphics[width=0.45\linewidth]{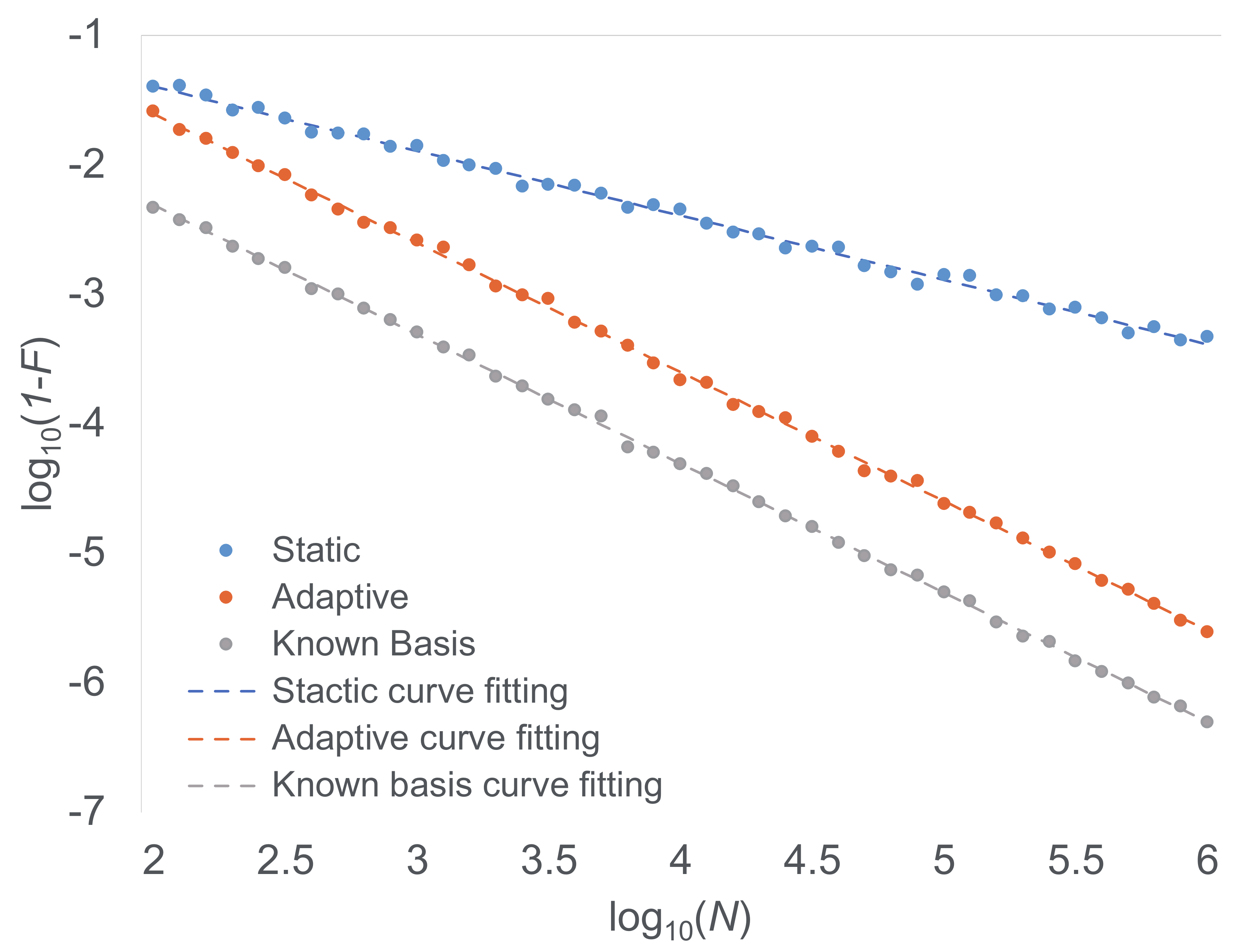}\label{fig:2:a}}\quad
\subfigure[]{\includegraphics[width=0.45\linewidth]{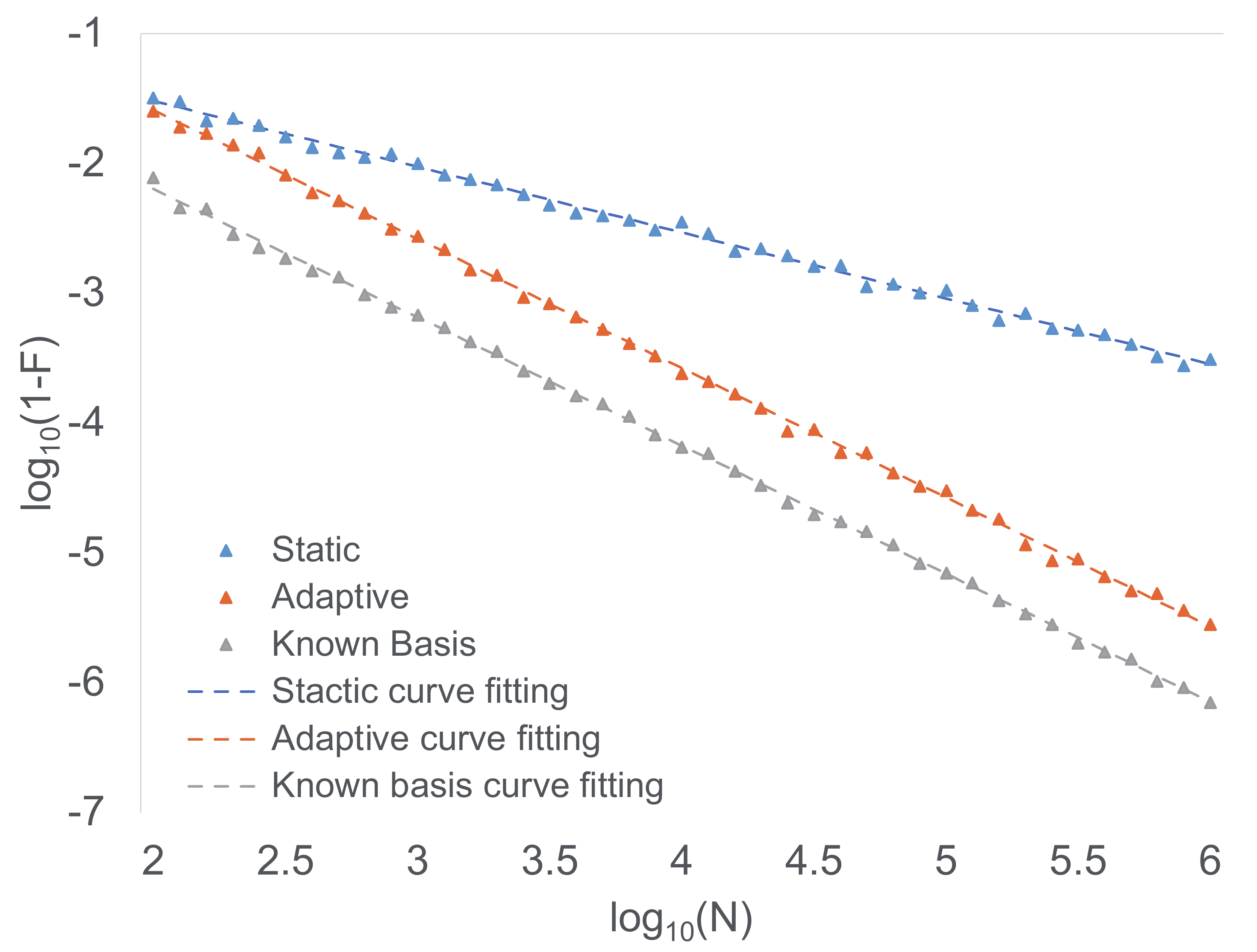}\label{fig:2:b}}
\addtocounter{figure}{-1}
 \caption{\label{fig:2}Monte Carlo simulations of different tomographic procedures for pure states: static tomography (blue), adaptive tomography (red) and known-basis tomography (grey). Every data point the infidelity averaged over 200 repetitions, and the dashed line represents corresponding power-law fitting to illustrate the scaling of infidelity versus $N$. (a) The results for tomography using SIC-POVM (for the state $\rho_1^\text{pure}$ in Eq. (\ref{eq:pure1})); (b) The results for tomography using MUB (for the state $\rho_2^\text{pure}$ in Eq. (\ref{eq:pure2})).}

\end{figure*}

\begin{figure*}
\centering
\addtocounter{figure}{1}
\subfigure[]{\includegraphics[width=0.45\linewidth]{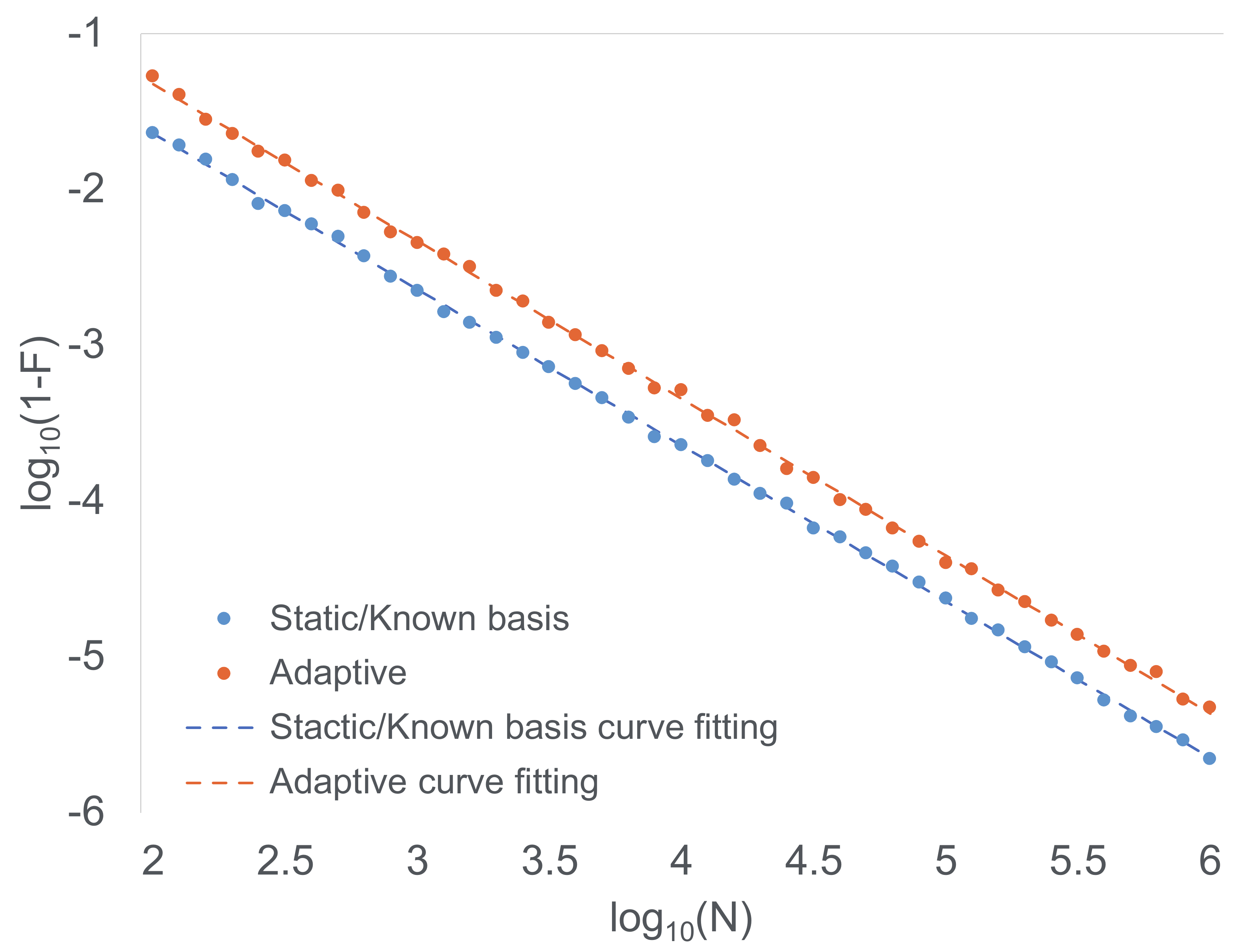}\label{fig:3:a}}\quad
\subfigure[]{\includegraphics[width=0.45\linewidth]{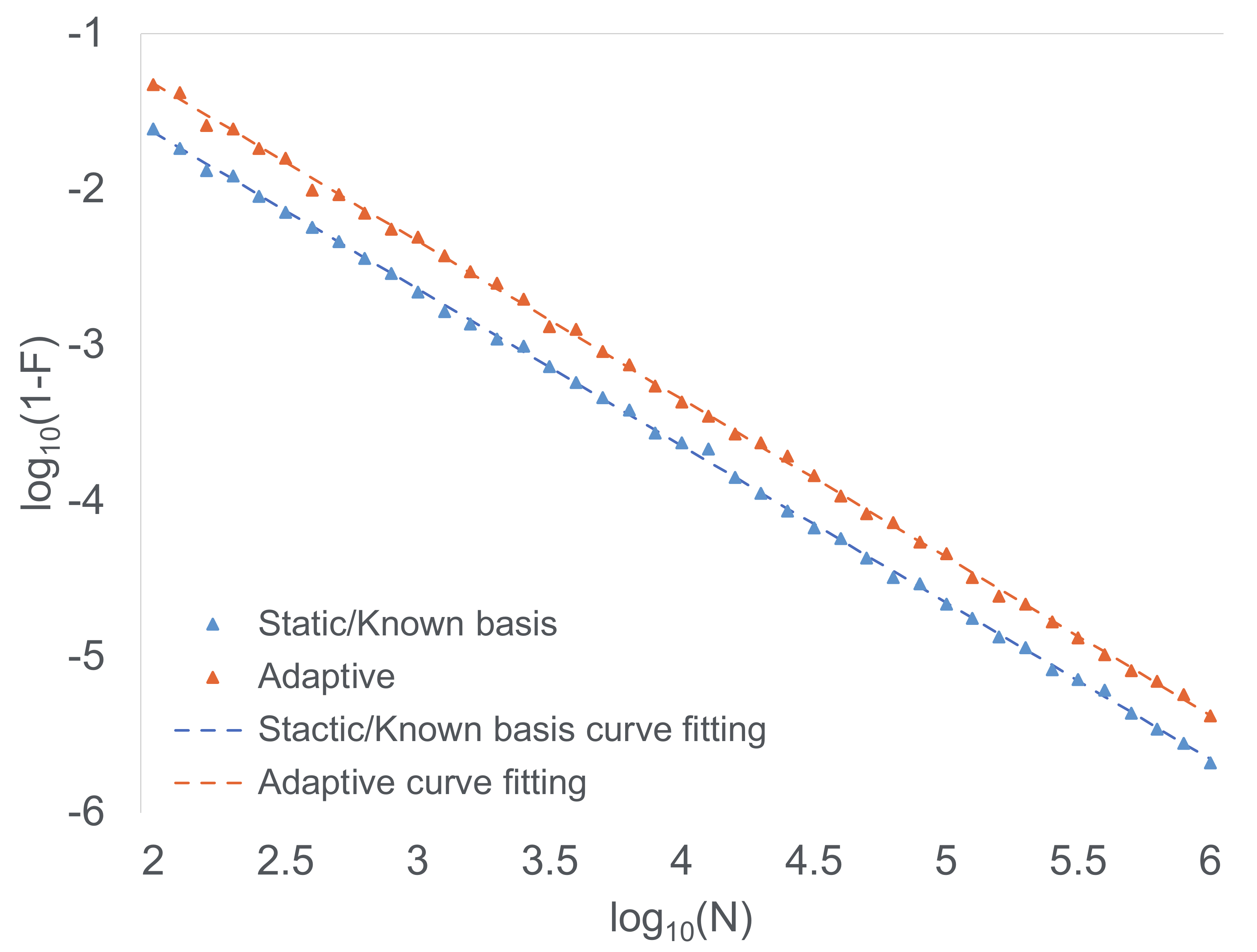}\label{fig:3:b}}
\addtocounter{figure}{-1}
 \caption{\label{fig:3}Monte Carlo simulations of different tomographic procedures for highly mixed states (the maximum mixed state $\rho^\text{mix}=\frac{1}{2}\openone$): static tomography/adaptive tomography (red) and known basis tomography (grey). (a) The results for tomography using SIC-POVM; (b) The results for tomography using MUB.}
\end{figure*}
\par
\textit{Simulation results}. We perform Monte Carlo simulations of the adaptive tomography for the pure state 
\begin{equation}
\rho_1^\text{pure} =
\begin{pmatrix}
1&0 \\
0&0
\end{pmatrix}
\label{eq:pure1}
\end{equation}
with initial outcome probabilities $\vec{p}=(1/2,1/6,1/6,1/6)$ for SIC-POVM; and 
\begin{equation}
\rho_2^\text{pure} =
\begin{pmatrix}
\frac{1}{2}+\frac{1}{\sqrt{3}}&\frac{1}{\sqrt{3}}-\frac{1}{\sqrt{3}}i \\
\frac{1}{\sqrt{3}}+\frac{1}{\sqrt{3}}i&\frac{1}{2}-\frac{1}{\sqrt{3}}
\end{pmatrix},
\label{eq:pure2}
\end{equation}
i.e., $\vec{s}=(1/\sqrt{3},1/\sqrt{3},1/\sqrt{3})$ for MUB. The results of static tomography and \textit{known-basis tomography} (knowing the true state and resorting to optimal measurement sets) are performed for comparison, as shown in Fig. \ref{fig:2}. Every marker represents an infidelity averaged over 200 repeated trials; the dashed line represents a linear fitting of $\log_{10}⁡(1-F)$ versus $\log_{10}⁡(N)$, which gives a relation $1-F=cN^\alpha$. The fitting results of $\alpha$ are summarized in Table. \ref{tab:fitting}. Notably, the adaptive protocol outperforms static tomography and approaches the scaling of known-basis tomography, for pure states. Figure \ref{fig:2} also shows the enormous advantages of the optimal strategy (known-basis tomography), departing from the typical static tomography. We obtain $c=0.4810$ and $c=0.5987$ in known-basis power-law fitting for SIC-POVM and MUB respectively, which is consistent with our predictions in Eqns. (\ref{eq:22}) and (\ref{eq:MUBinf}).
\begin{table}
 \centering \caption{\label{tab:fitting}Curve fitting results of $\alpha$ for pure states.}
\begin{ruledtabular}
\begin{tabular}{ccc}
    &\textbf{SIC-POVM}&\textbf{MUB}\\
    \hline
    Static & $-0.4977\pm0.0054$ & $-0.5069\pm0.0051$\\
    Adaptive & $-0.9976\pm0.0042$ & $-0.9951\pm0.0048$\\
    Known-basis & $-0.9979\pm0.0036$ & $-0.9873\pm0.0043$
   \end{tabular}
\end{ruledtabular}
    \end{table}
\par
In the same way, we perform simulations for the maximum mixed state $\rho^\text{mix}=
\begin{pmatrix}
1/2&0 \\
0&1/2
\end{pmatrix}$, as shown in Fig. \ref{fig:3}. In this case, the static tomography and known-basis tomography are identical. All the three procedures scale as $O(1/N)$; the adaptive tomography shows no superiority compared with the non-adaptive one.
\par
In summary, due to the physical constraint on the reconstructed state, different configurations for pure states will result in distinct performance of tomography, as we derived in Sec. \ref{sec:accuracy}. Therefore, the adaptive protocol can improve the accuracy significantly for pure states. Yet for highly mixed states, the measurement results hardly violate the physical constraint. As a result, the accuracy of tomography is completely determined by the statistical fluctuations.
\par
\begin{figure}
    \includegraphics[width=\linewidth]{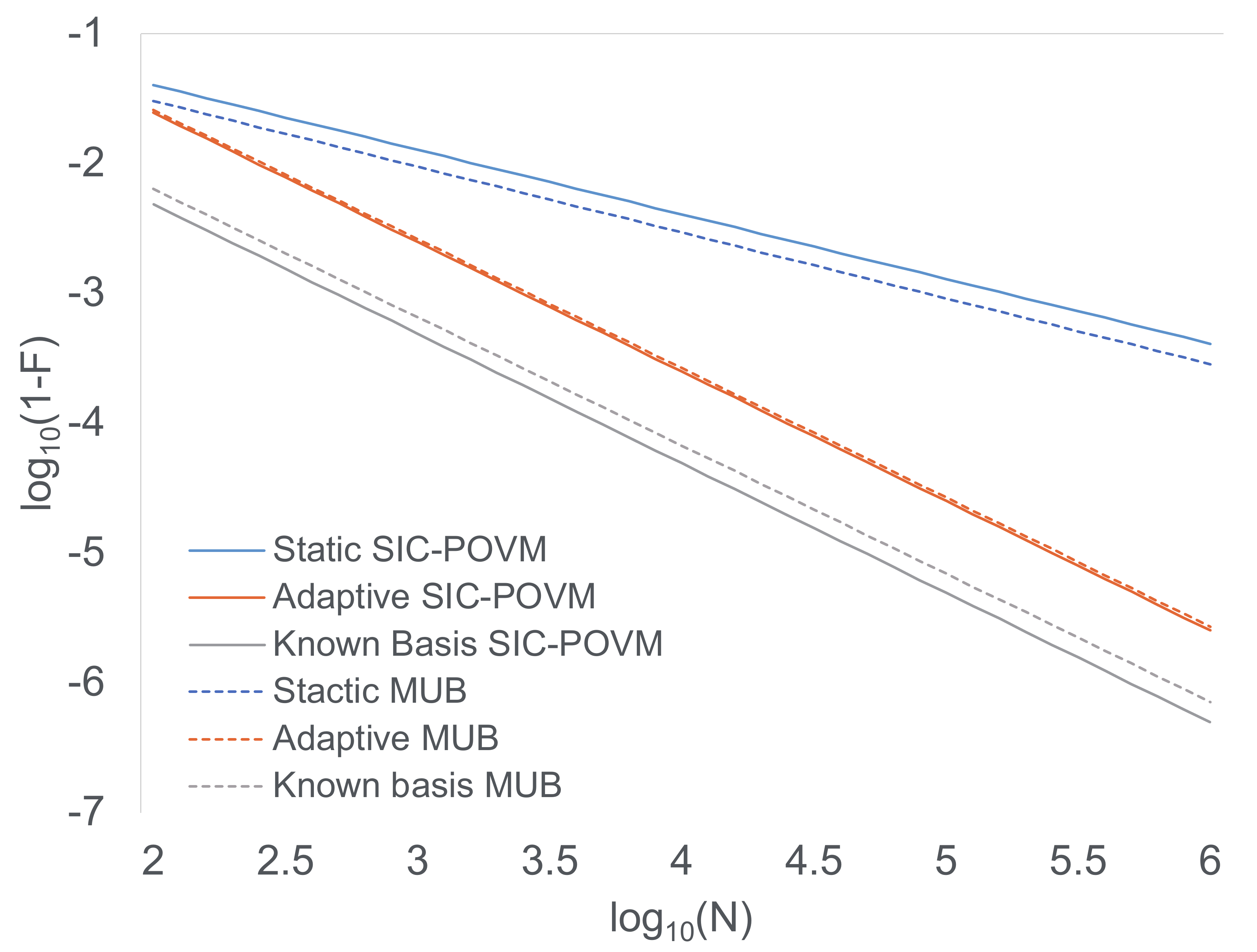}
  \caption{\label{fig:compare1}The comparison of the performance of MUB (dashed lines) and SIC-POVM (solid lines) in different tomographic procedures for pure states: static tomography (blue), adaptive tomography (red) and known-basis tomography (grey).}
\end{figure}

Figure \ref{fig:compare1} compares the fitting curves of tomography using SIC-POVM and MUB. MUB reveals better accuracy in static tomography, while the SIC-POVM performs better in optimal configuration. In the two-step adaptive procedure, how close the configuration in the second step is to the optimal configuration depends on the accuracy of the first step, which will in turn affect the accuracy of the second step. Consequently, the accuracy of the final estimation depends on both the statistical fluctuations in the first step and the accuracy of the optimal configuration. The better accuracy of MUB in pre-estimation results in a smaller discrepancy for the optimal configuration in the second step, where SIC-POVM obtains more accurate estimation. The estimation accuracy in the two stages for MUB ($3/2N$ for typical static tomography and $2/3N$ for known-basis tomography) and SIC-POVM ($2/N$  for typical static tomography and $1/2N$  for known-basis tomography) are exactly canceled out. Finally, MUB and SIC-POVM reveal similar accuracy in adaptive tomography.

\section{\label{sec:4}Nearly-pure states in quantum tomography}
\par
The adaptive protocol manifests its superiority for pure states, but not for highly mixed states. As far as we know, nearly-pure states with a moderate mixture are more popular cases in physical implementations of QIP, therefore understanding the performance of adaptive tomography for nearly-pure states is a key for its applications in practical setups. In the first instance, we perform simulations for states of the form $\rho_1^\text{near} =(1-\lambda)\rho_1^\text{pure}+\lambda\openone$ [$\rho_2^\text{near} =(1-\lambda)\rho_2^\text{pure}+\lambda\openone$ for MUB], where $\lambda$ is the smaller eigenvalue of $\rho_1^\text{near}$ ($\rho_2^\text{near}$). As an example, the simulation result for $\lambda=0.0002$ using SIC-POVM is shown in Fig. \ref{fig:nearsic}, which reveals remarkable distinction from pure states. It is shown in Sec. \ref{sec:mle} that MLE demonstrates its superiority for completely pure states; conversely, statistical fluctuations are mainly responsible for the estimation of highly mixed states. In the transition region of the two kind of states, the effects of purity and statistical fluctuations are comparable and the mechanism of tomography still remains unclear, even for single qubit.
\begin{figure}
    \includegraphics[width=\linewidth]{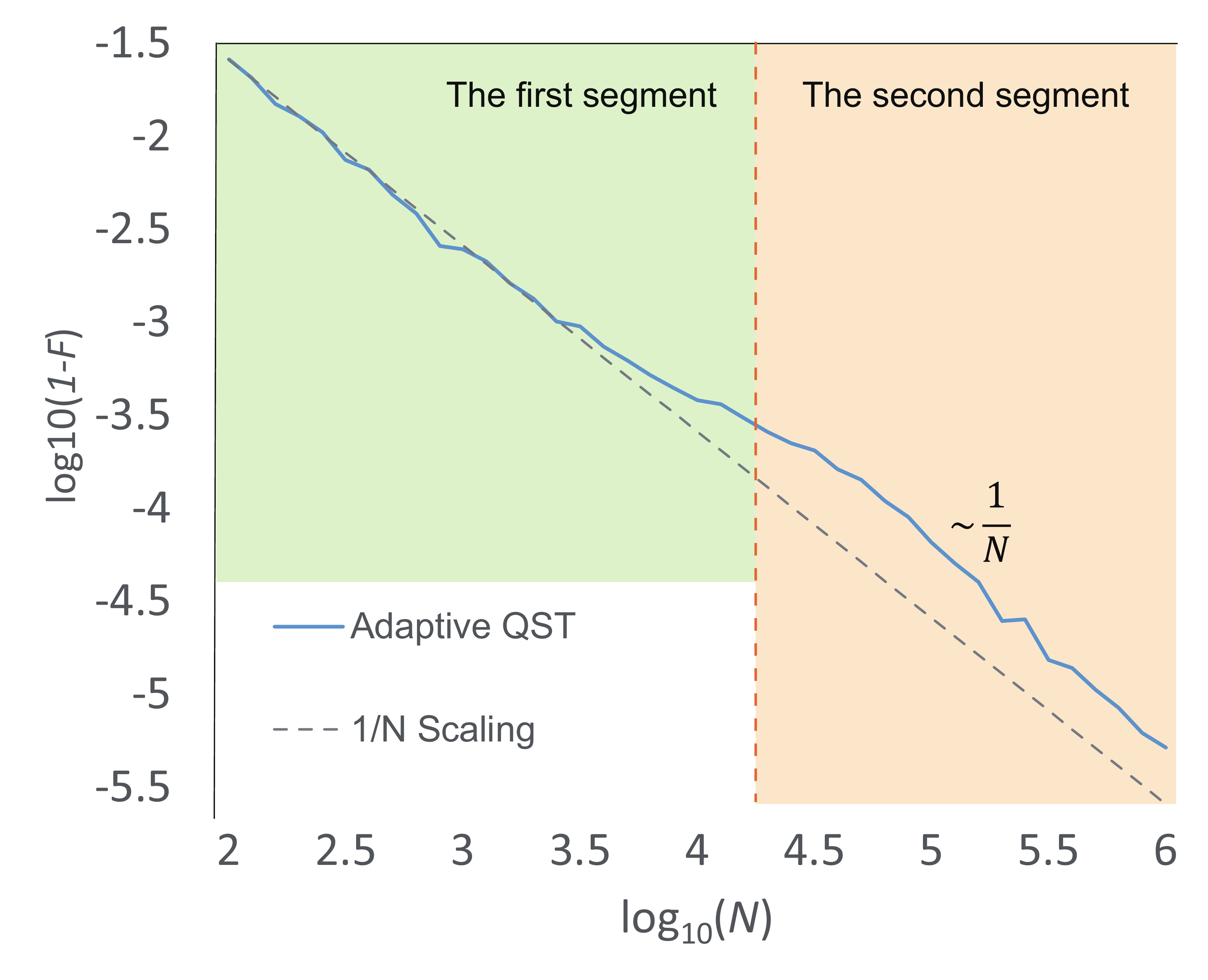}
  \caption{\label{fig:nearsic}Monte Carlo simulations of adaptive tomography using SIC-POVM. The true state is a nearly pure state $\rho_1^\text{near}=
\begin{pmatrix}
0.9998&0 \\
0&0.0002
\end{pmatrix}$. The curve contains two segments and a turning point. The scaling of the infidelity becomes ``\textit{invalid}'' around the demarcation of two segments (vertical dashed line).}
\end{figure}

\par
In practice, we simulate a range of cases and observe that the curve contains a turning point, which divides the curve into two segments. The adaptive tomography achieves $O(1/N)$ scaling at large $N$, but becomes ``\textit{invalid}'' at small $N$ ($N<O(1/\lambda)$ in the simulation). Notably, the ``invalidation'' for nearly-pure states also exists in tomography based on iterative algorithms \cite{RN14}. To date, there has not been a theory to rigorously interpret this phenomenon. 

\subsection{\label{sec:nearly}Nearly-pure states in the optimal configuration}
\par
We start from the analysis of the first segment in Fig. \ref{fig:nearsic}. As we show in Appendix, the ``invalidation'' at small $N$ is attributed to the statistical fluctuations of the outcomes instead of the impurity of the state. When the probability $p_i$ of observing one of outcomes $i$ is exceedingly small, a small $N$ is insufficient to estimate the parameter $p_i$. Envisage a scenario which determines a parameter $p_i=1/10000$ utilizing samples $N=100$ within one trial. For each trial, we may obtain $\hat{p}_i=0.01$ with a probability 0.01 and $\hat{p}_i=0.01$ wih a probability 0.99. If we repeat the experiment with many trials, the expectation of $\hat{p}_i$ should be 1/10000. This situation is exactly what we encounter in QST for nearly-pure states, resulting in a decrease of the average fidelity.
\begin{figure}
\centering
\addtocounter{figure}{1}
\subfigure[]{\includegraphics[width=0.45\linewidth]{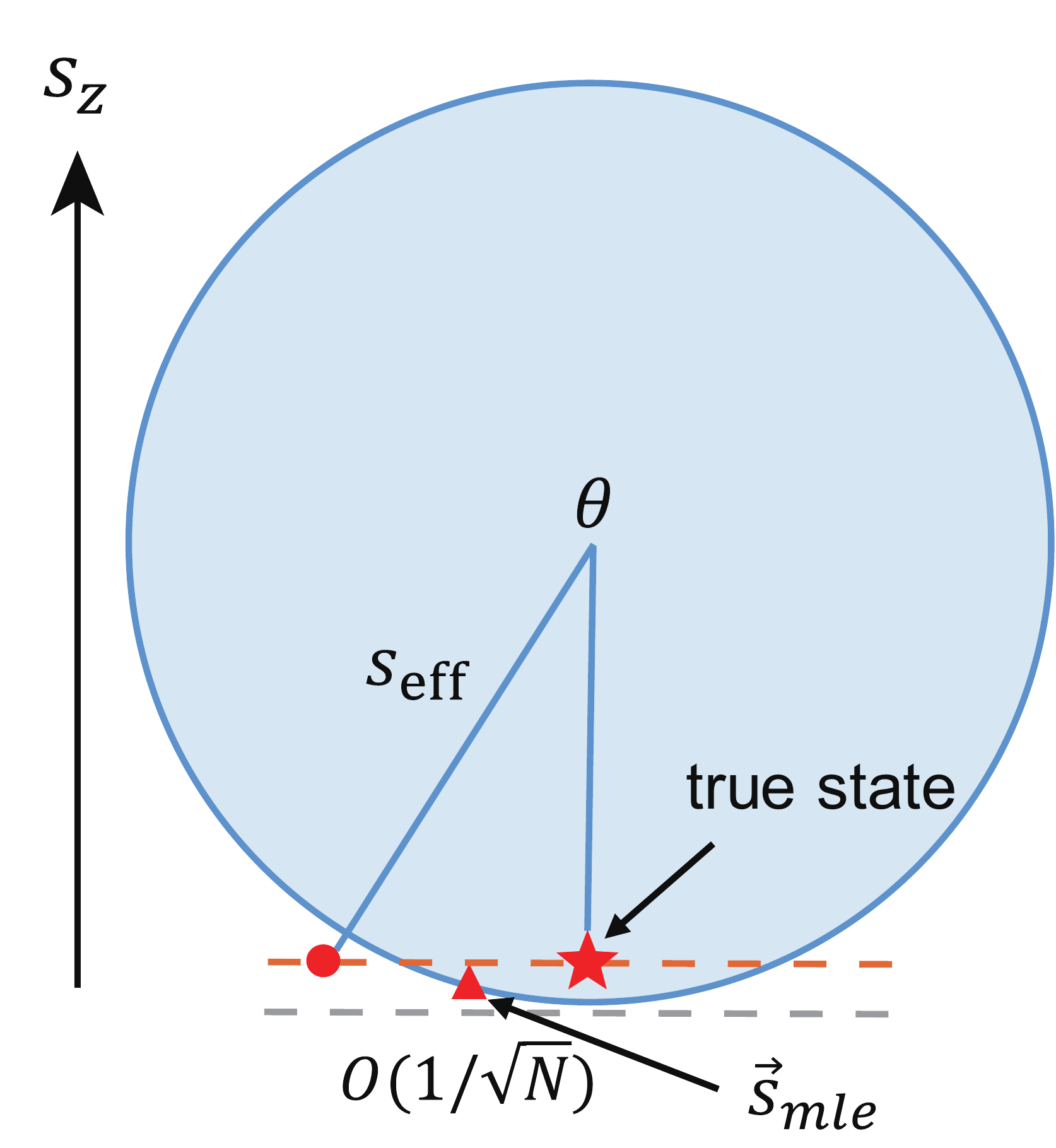}\label{fig:6:a}}\quad
\subfigure[]{\includegraphics[width=0.45\linewidth]{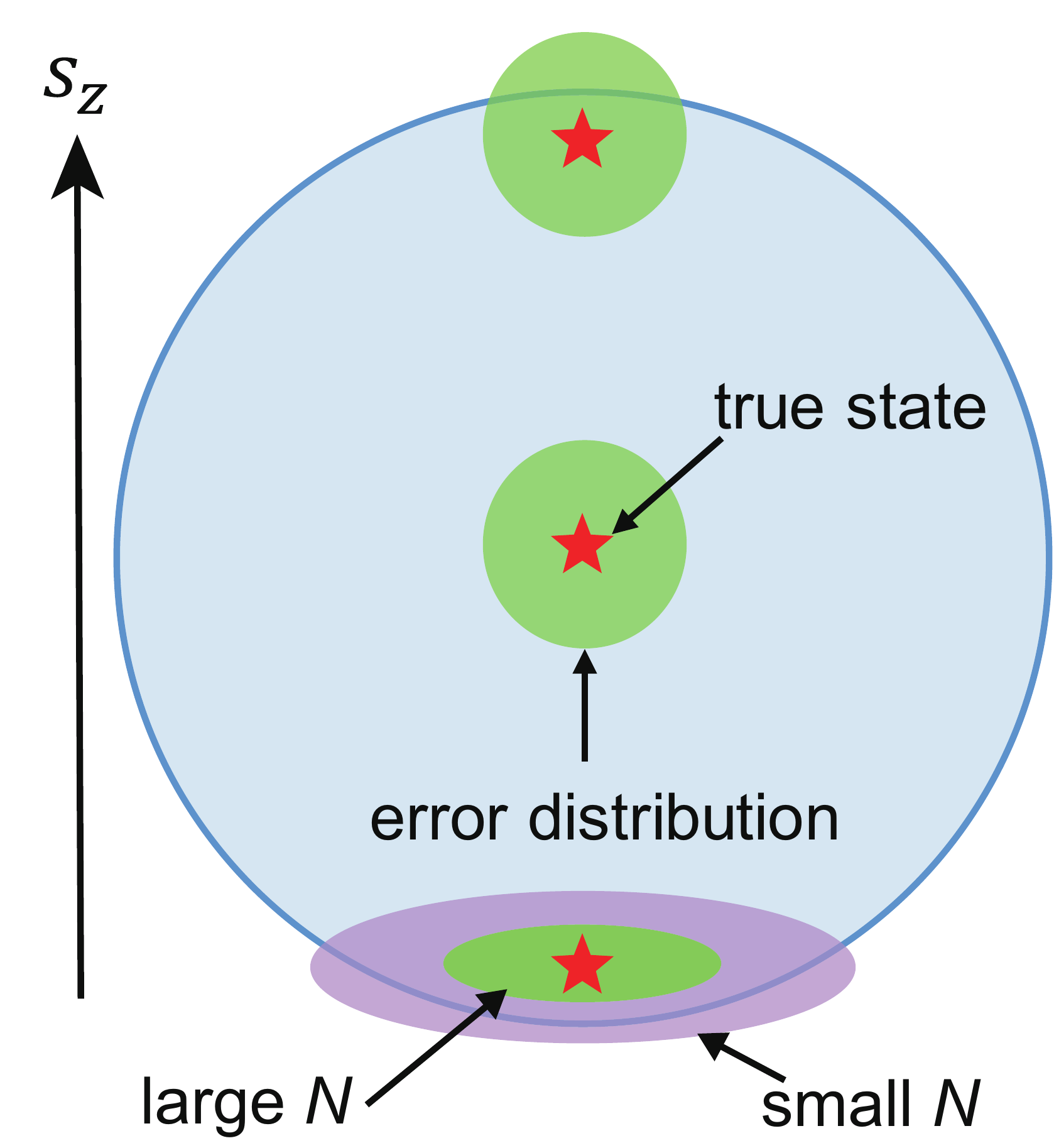}\label{fig:6:b}}
\addtocounter{figure}{-1}
 \caption{\label{fig:6}The cross-section of Bloch sphere to illustrate the adaptive quantum state tomography. (a) State estimation for a nearly-pure state (red star) in the optimal configuration. The tangent plane of Bloch sphere (dashed line, grey) denotes the ensemble of measurement results with a high probability to obtain. The representative measurement results (red dot) are inferred by $\vec{s}_\mathrm{eff}$ with $\theta\sim O(1/\sqrt{N})$, and $\vec{s}_\mathrm{mle}$ denotes the MLE of $\vec{s}_\mathrm{eff}$. (b) The distribution of the estimation $\hat{\rho}$ (shaded area) for tomography using SIC-POVM. The distribution of $\hat{\rho}$ (shaded area) depends on both the purity of the true state (red star) and the number of samples $N$.}
\end{figure}

\par
Figure \ref{fig:6:a} illustrates the situation of estimating nearly-pure states with the optimal configuration. For $N\ll 1/\lambda$, it is highly probable to obtain a measurement result in the tangent plane of Bloch sphere, and MLE will obtain a pure state inferred as $\vec{s}_\mathrm{mle}$. This situation is very close to the estimation of a pure state in the optimal configuration, therefore we expect that the infidelity can be estimated through $\vec{s}_\mathrm{mle}$ in the first segment in Fig. \ref{fig:nearsic}. With the increase of $N$, the statistical fluctuations decrease and $s_\mathrm{eff}\leq 1$. The measurement results are not likely to violate the positivity constraint. This situation is very close to the estimation of a highly mixed state, therefore the infidelity can be estimated by Eqns. (\ref{eq:10}) and (\ref{eq:11}) directly in the second segment. Recall that in Eq. (\ref{eq:eff}) we introduce $s_\mathrm{eff}$ to denote the typical length of the reconstructed Pauli vector. $s_\mathrm{eff}$ can be considered as kind of figure of merit which represents how close the measurement result is from a pure state. Based on the above analysis, we can set $s_\mathrm{eff}=1$ as the turning point of the two segments, then the average infidelity can be evaluated by
\begin{equation}
1-F=
\begin{dcases}
\frac{1}{2}\left(1-\vec{s}\cdot\vec{s}_\mathrm{est}\right) & s_\mathrm{eff}> 1 \\
\frac{1}{2}\left(1-\vec{s}\cdot\vec{s}_{\mathrm{est}}-\sqrt{1-s^2}\sqrt{1-{s_{\mathrm{est}}}^2}\right) & s_\mathrm{eff}\leq 1
\end{dcases}
\label{eq:26}
\end{equation}
\par
For nearly-pure states in optimal configuration, the criterion $s_\mathrm{eff}\leq 1$ could be simplified approximately as $N\geq2/\lambda$, where $\lambda=(1-s)/2$. With Eq. (\ref{eq:26}), one can conclude
\par
(i) For SIC-POVM:
\begin{align}
1-F&=\frac{1}{2}\left(1-s s_\mathrm{est}\cos{\theta}\right) \nonumber \\
&\approx\frac{1}{2}\left(1-s\right)+\frac{1}{4}\theta^2 s\approx\lambda+\frac{1}{2N} \  &N\ll\frac{2}{\lambda};\label{eq:27}\\
1-F&\approx\frac{1}{4}{|\delta\vec{s}|}^2\approx\frac{2}{N} \quad &N\gg\frac{2}{\lambda}.
\label{eq:28}
\end{align}
\par
(ii) For MUB:
\begin{align}
1-F\approx&\lambda+\frac{2}{3N} \quad& N\ll&\frac{3}{2\lambda},  \label{eq:MUBinf1}\\
1-F\approx&\frac{3}{2N} \quad& N\gg&\frac{3}{2\lambda}.
\label{eq:MUBinf2}
\end{align}
The theoretical predictions from Eq. (\ref{eq:27}) and Eq. (\ref{eq:MUBinf1}), together with the $1/N$ scaling for the second segment, are shown in Fig. \ref{fig:infnear} (dashed line, grey). Simulation results (grey data) are in well agreement with our theoretical predictions. Note the deviation from theoretical prediction is observed near the turning point, because the contribution of MLE attenuates gradually with the increase of $N$. Nonetheless, our theoretical model is sufficient to give clear description of the tomography accuracy and is in well agreement with the turning point.

\subsection{Adaptive tomography}
\par
Let us proceed to deduce the infidelity for adaptive protocol. The pre-estimation step gives a preliminary knowledge of the state with typical error as $O(1/\sqrt{N_1})$. This statistical fluctuation introduces a discrepancy from the optimal configuration in the second step. On average, this discrepancy is equivalent to an extra $O(1/N_1)$ impurity in the final estimation, and can be encapsulated in a coefficient $\beta$ for SIC-POVM ($\gamma$ for MUB). Recall that the pre-estimation step consume $N_1=N/2$ samples of resources. Therefore, two modifications should be applied on Eqns. (\ref{eq:27}) and (\ref{eq:28}) in adaptive tomography: (i) $\lambda$ should be replaced by $\lambda+\beta/N_1$ for SIC-POVM($\lambda+\gamma/N_1$ for MUB); (ii) $N$ should be replaced by $N/2$. Then the formulas of infidelity can be recast into
\par
(i) For SIC-POVM:
\begin{align}
1-F\approx&\lambda+\frac{2\beta}{N}+\frac{1}{N} \quad& N\ll&\frac{4}{\lambda},  \label{eq:29}\\
1-F\approx&\frac{4}{N} \quad& N\gg&\frac{4}{\lambda}.
\label{eq:30}
\end{align}
\par
(ii) For MUB:
\begin{align}
1-F\approx&\lambda+\frac{2\gamma}{N}+\frac{4}{3N} \quad& N\ll&\frac{3}{\lambda},  \label{eq:AMUB1}\\
1-F\approx&\frac{3}{N} \quad& N\gg&\frac{3}{\lambda}.
\label{eq:AMUB2}
\end{align}
\begin{figure*}
\centering
\addtocounter{figure}{1}
\subfigure[]{\includegraphics[width=0.45\linewidth]{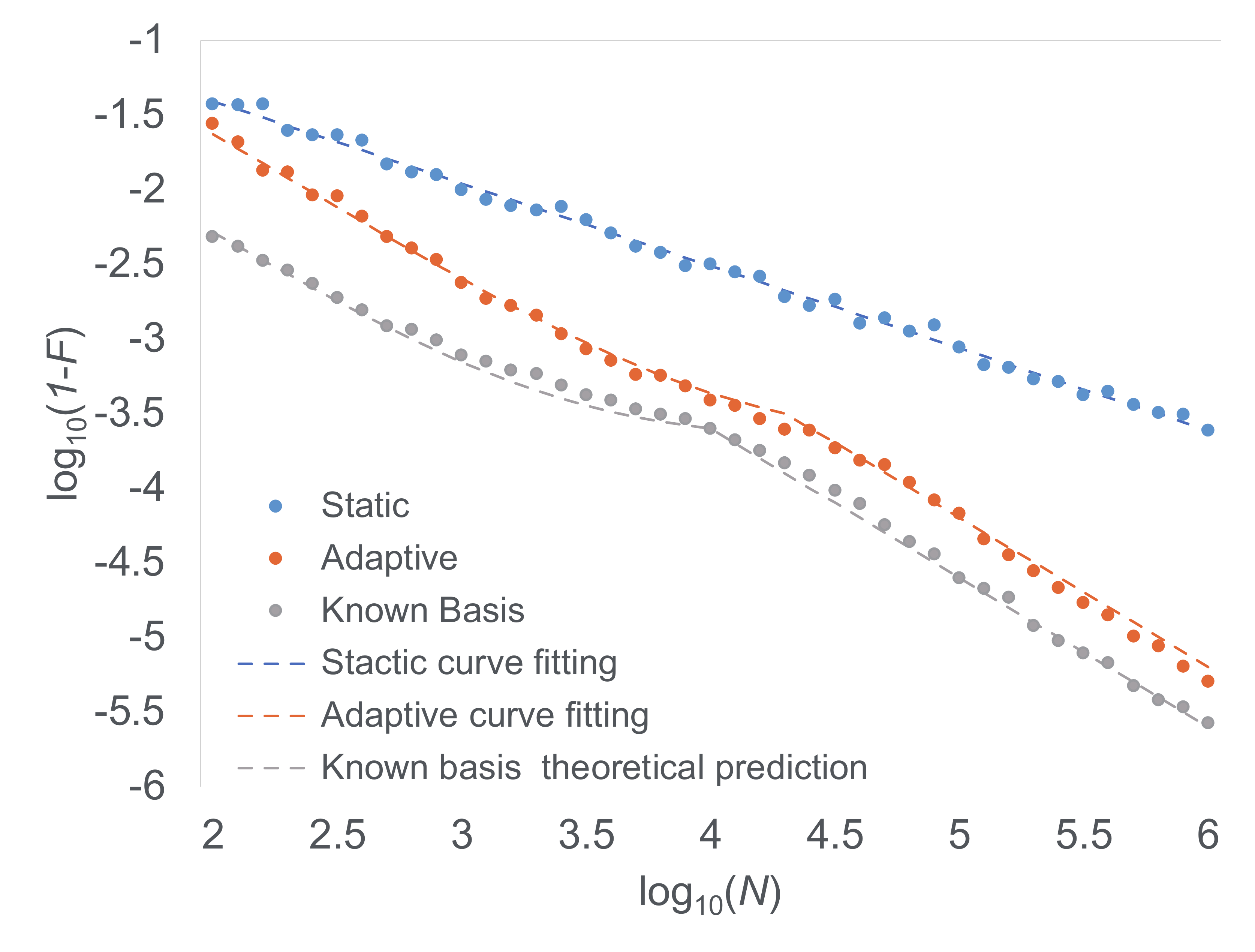}\label{fig:infnear:a}}\quad
\subfigure[]{\includegraphics[width=0.45\linewidth]{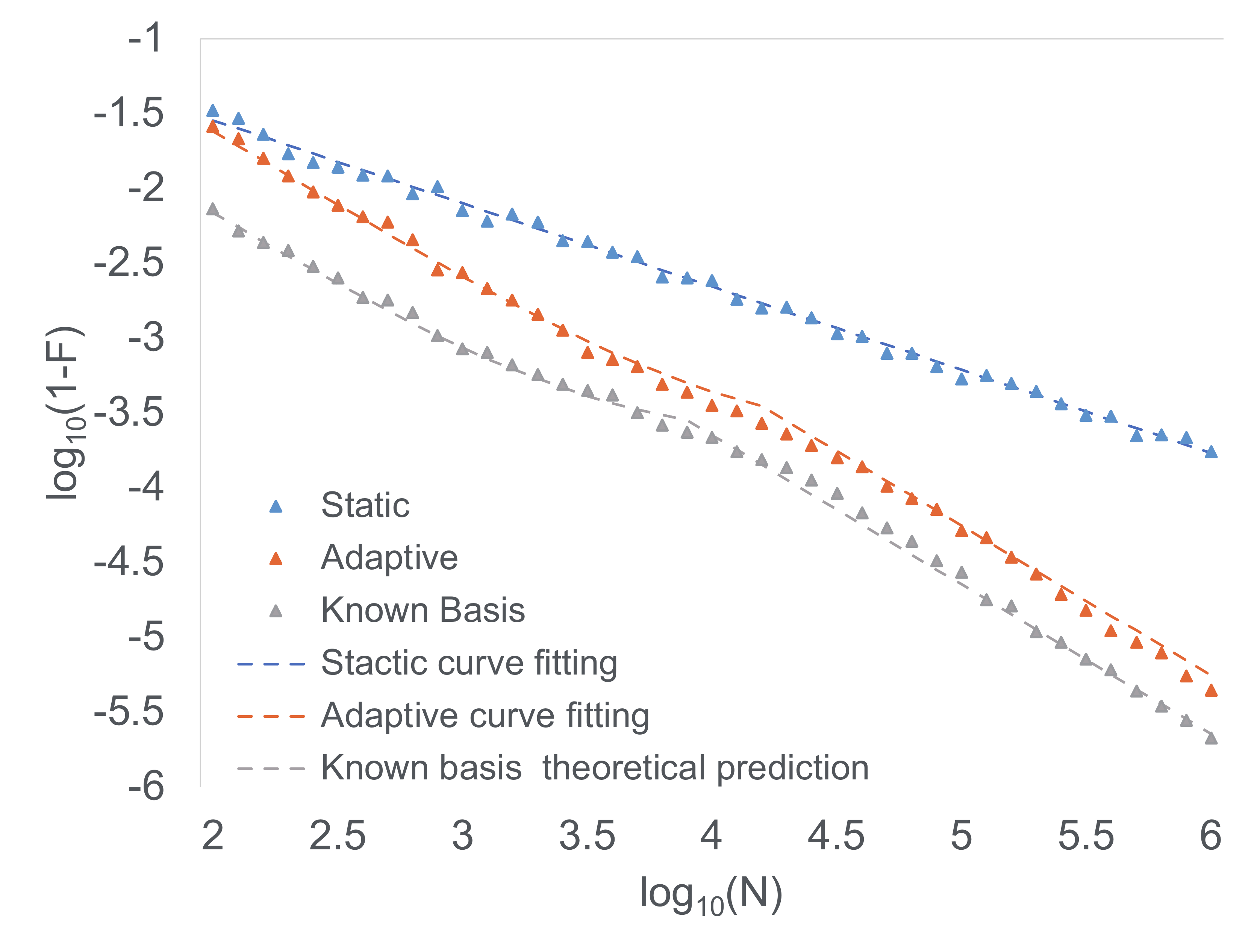}\label{fig:infnear:b}}
\addtocounter{figure}{-1}
 \caption{\label{fig:infnear}Tomography results for the nearly-pure state with a small eigenvalue $\lambda=0.0002$ using two measurements: (a) SIC-POVM (for the state $\rho_1^\text{near}$); (b) MUB (for the state $\rho_2^\text{near}$). Every marker represents a infidelity averaged over 200 repeated trails. Blue: static tomography. Red: adaptive tomography. Grey: known-basis tomography.}
\end{figure*}
\par
Figure \ref{fig:infnear} shows the results for the nearly-pure state $\rho$ using static tomography, adaptive tomography and known basis tomography respectively. The adaptive scheme significantly improves the average fidelity, and achieves $1/N$ scaling for large $N$. For modest number of samples, adaptive tomography of nearly-pure states performs more prominent improvements on accuracy in comparison with pure states, and approaches the fidelity of known-basis tomography, especially at the number $N$ around the turning point. The dependence of the infidelity on $N$ reveals how the purity of the state and the statistical fluctuations affect the accuracy, as we discussed in Sec. \ref{sec:nearly}.

\begin{figure}
    \includegraphics[width=\linewidth]{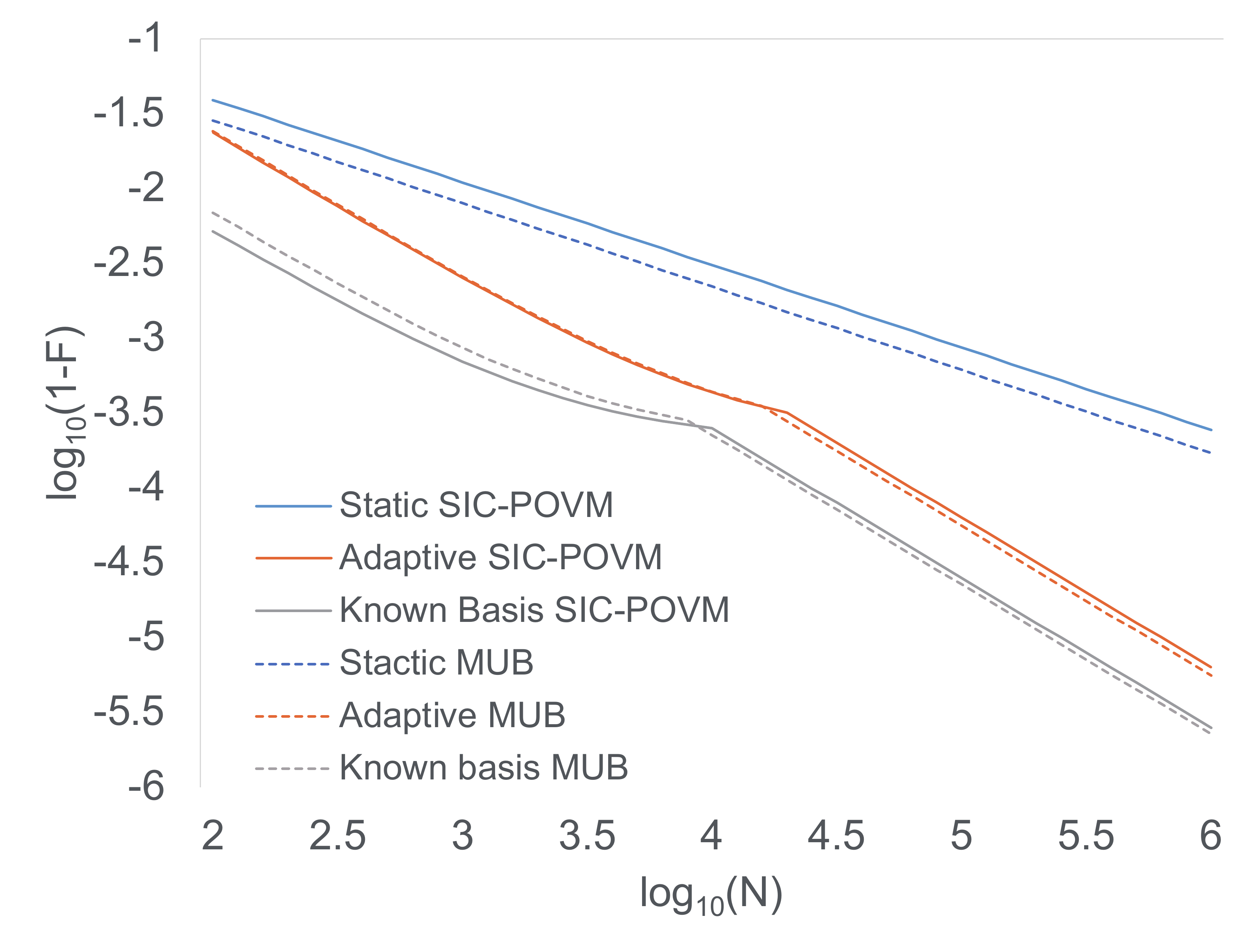}
  \caption{\label{fig:compare2}The comparison of the performance of MUB (dashed lines) and SIC-POVM (solid lines) in different tomographic procedures for nearly-pure states: static tomography (blue), adaptive tomography (red) and known-basis tomography (grey).}
\end{figure}
\par
By comparing SIC-POVM and MUB, we conclude that MUB still performs better in static tomography for nearly-pure states. In known-basis tomography, SIC-POVM shows better fidelity at small $N$, and behaves similarly as MUB at large $N$. The performance of adaptive tomography is similar for the two measurements. The similar performance is also testified by the fitting results $\beta=0.6662\pm 0.0292$ and $\gamma=0.5296\pm 0.0362$, which lead to very close values for Eq. (\ref{eq:29}) and Eq. (\ref{eq:AMUB1}).

\subsection{Criterion of nearly-pure states}
\par
Back from infidelity to Bloch sphere, we now explain what ``nearly-pure states'' is in terms of QST. In previous works, the criteria of nearly-pure states was considered as $\lambda <O(1/\sqrt{N})$. It has been discussed in \cite{RN12} that why adaptive tomography achieves a better infidelity. Standard tomography does not minimize the expected infidelity for two reasons: first, the variance of the estimate $\hat{\rho}$ depends also on $\rho$ itself; second, the dependence of infidelity on the error, $\Delta\rho$, also varies with $\rho$. Ulteriorly, we assert that criterion of nearly-pure states is also \textit{anisotropic} and \textit{inhomogeneous}, varies with $\rho$ , not invariably $\lambda<O(1/\sqrt{N})$.
\par
It is reasonable that a state is ``nearly-pure'' when the reconstructed state $\hat{\rho}$, the linear inversion of the measurement results, can be a pure state. In other words, if the error distribution of measurement results is all within the Bloch sphere, the state behaves as a complete mixed state, therefore the scaling of infidelity turn to $O(1/N)$. For instance, in the optimal configuration for SIC-POVM, the distribution of $\hat{p}_0$ is defined as Eq. (\ref{eq:13}). The state becomes nearly-pure when the standard deviation $\Delta\hat{p}_0$ exceeds $p_0$ itself:
\begin{equation}
\sqrt{\frac{1}{N}}\sqrt{p_0(1-p_0)}>p_0.
\label{eq:31}
\end{equation}
The result $\lambda<2/(N+1)$ is in accordance with our simulation results and theoretical derivation (Fig. \ref{fig:nearsic}). One may wonder the question: why the turning point exists? The turning point is existent for all mixed states, typically at $\lambda\sim 1/\sqrt{N}$, but not always at $1/\sqrt{N}$, especially when the optimal configuration is applied. As shown in Fig. \ref{fig:6:b}, the error is compressed in some directions in the Bloch sphere. The purity of the state and the statistical fluctuations of the measurement outcomes jointly define the transition region in the performance of the adaptive tomography. The optimal configuration actually shifts the turning point, rather than transfers the scaling from $O(1/\sqrt{N})$ to $O(1/N)$. Our results are generally valid for all mixed states, and become visible for nearly-pure states, rather than highly mixed states. Pure states are special cases that the smaller eigenvalue is zero.
\par
In adaptive tomography, diagonalizing the density matrix $\rho$ is not the recipe of achieving a better fidelity, but the probability $p_i$ of observing one of the POVM outcomes equals (or approaches) 0 is crucial. Both parallel and antiparallel strategies in SIC-POVM diagonalize the density matrix but provide totally different performance on nearly pure states. In substance, the recipe is to eliminate (or reduce) the \textit{randomness} of outcomes. In optimal configuration for pure states, the error in one of the outcomes is eliminated completely, rendering the response of this outcome from a \textit{probabilistic} process to a \textit{deterministic} one. Therefore, absolute $1/N$ scaling would be achieved if the state is completely pure and the measurement is aligned in the truly optimal configuration, when the full information about the state is known.

\section{\label{sec:5}Conclusion}
\par
In conclusion, we investigated adaptive quantum state tomography using mutually unbiased bases (MUB) and symmetric informationally complete positive operator-valued measure (SIC-POVM), both of which give rise to $O(1/N)$ infidelity for large $N$. In comparison with previous works, we gave more detailed derivation and clear picture for scaling improvement of adaptive strategy compared to the conventional tomography. We further interpreted why adaptation achieves a better fidelity by considering the effect of stochastic fluctuations. In particular, we discussed how the interplay between the purity of the state and the stochastic fluctuations of measurement results affect the accuracy of reconstruction. Our analyses can be generalized to the tomography of quantum states with more than a single qubit. Specifically, the effects of the purity of the state on the accuracy of reconstruction would be equally effective. Our results highlight the unique behavior of nearly-pure states in quantum state tomography. 
\appendix*
 
\section{Statistical fluctuations and impurity for nearly-pure states}
\par
In the scenario of QST, there are two differences for the estimated state between a pure state and a nearly-pure state: (i) the probabilities of the outcomes $p_i$, which is directly relevant to the statistical fluctuations; (ii) the purity of the true state, which is quantified by the eigenvalues of the states. To identify why the adaptive protocol becomes ``\textit{invalid}'' for nearly-pure states, we perform static tomography using SIC-POVM for two states: the misaligned pure state $\rho_1$ (by which we refer to the misalignment from the optimal configuration) and the nearly pure state $\rho_2$:
$$\rho_1={\begin{pmatrix}
0.0002 & 0.0141i\\
0.0141i & 0.9998 
\end{pmatrix}},\rho_2={\begin{pmatrix}
0.0002 & 0\\
0 & 0.9998
\end{pmatrix}}.$$
\begin{figure*}
\centering
\addtocounter{figure}{1}
\subfigure[]{\includegraphics[width=0.45\linewidth]{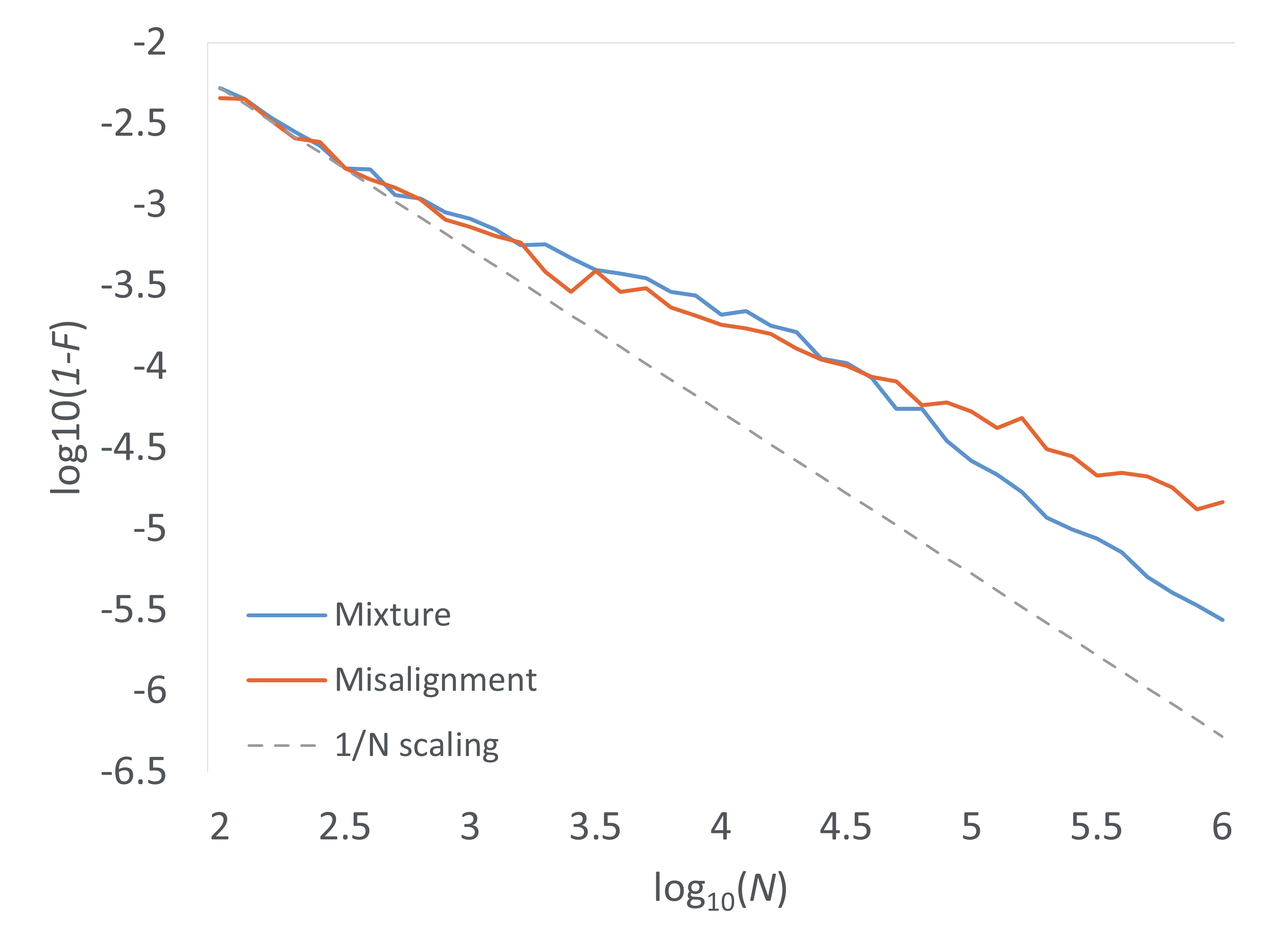}\label{fig:9:a}}
\subfigure[]{\includegraphics[width=0.45\linewidth]{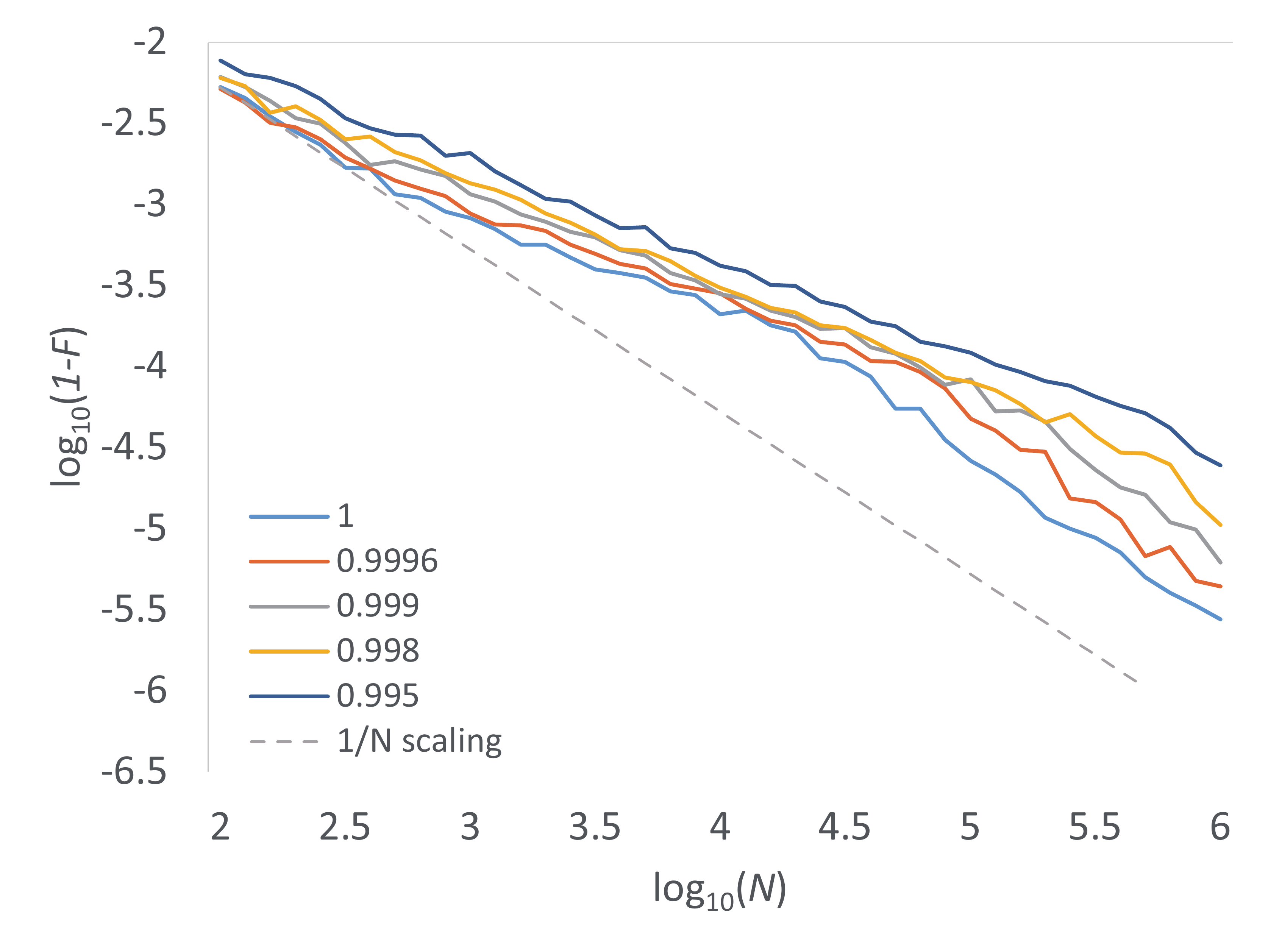}\label{fig:9:b}}
\addtocounter{figure}{-1}
\caption{\label{fig:9}Simulation results for static tomography using SIC-POVM: (a) the misaligned pure state $\rho_1$ (red) and the nearly pure state $\rho_2$ (blue). The two states perform similarly in the first segment, but diverges at higher $N$. (b) The states with same purity but various misalignments, which are defined by the discrepancy angle $\theta$ from the optimal configuration. The solid lines in different colors represent different misalignments with respect to the optimal configuration.}
\end{figure*}

Both of the two states lead to the same probability $p_0=0.0001$ for the outcome $0$, but only the nearly pure state $\rho_2$ has some impurity ($\lambda=0.0002$). The simulation results of static tomography for the two states are shown in Fig. \ref{fig:9:a}. We conclude from this comparison that the infidelities for both states behave similarly at small $N$, which should be attributed to the probability $p_0=0.0001$; conversely, the scaling of two states diverge at higher number of samples ($\log_{10}N>4.0$). The infidelity of the nearly-pure state $\rho_2$ transforms into $O(1/N)$ scaling while the infidelity of the misaligned pure state $\rho_1$ maintains $O(1/\sqrt{N})$ scaling.

\par
What if the states have the same purity but different misalignments? The results for static tomography are illustrated in Fig. \ref{fig:9:b}. The misalignment is quantified by its angle $\theta$ deviated from the optimal configuration. We take $\cos\theta=1,0.9996,0.999,0.998,0.995$ in the numerical simulations respectively. The turning point in each situation is obvious compared with the $1/N$ scaling. It infers that even the states have identical degree of impurity, these states reveal different manifestations for a range of misalignments, due to the fact that the statistical fluctuations are not the same, and result in different conditions of nearly-pure states.

\begin{acknowledgments}
This work was supported by the National Natural Science Foundation of China under Grants No. 61490711, No. 91536113 and No. 11474159. The authors thank Bei Zeng and Dawei Lu for helpful discussions. A. Zhang expresses his gratitude to Ya Xu for valuable comments on this manuscript. 
\end{acknowledgments}


%

\end{document}